\RequirePackage{ifpdf}
\ifpdf
	\documentclass[twocolumn, letterpaper, prb, showkeys, showpacs, pdftex]{revtex4}
\else
	\documentclass[twocolumn, letterpaper, prb, showkeys, showpacs, dvips]{revtex4}
\fi

\usepackage[nodvipsnames]{color}
\usepackage{subfigure,amsmath}

\ifpdf
  \usepackage[pdftex]{graphicx}
  \graphicspath{{pdf-figures/}}
\else
  \usepackage[dvips]{graphicx}
  \graphicspath{{eps-figures/}}
\fi

\usepackage[%
\ifpdf
	pdftex,%
	pdfstartview = {FitH},%
\else
    dvips,%
\fi
    pdftitle = {{Microstructure and mechanical properties of constrained shape-memory alloy nanograins and nanowires}},%
	pdfsubject = {{We use the phase-field method to study the martensitic transformation at the nanoscale, in particular the effect of size and geometry on the microstructure.}},%
	pdfauthor = {{Mathieu Bouville and Rajeev Ahluwalia}},%
	pdfkeywords = {{nanomaterials, phase-field simulations, shape-memory material, nanowires}},%
    hyperindex,%
    bookmarksopen,%
    bookmarksopenlevel=1,%
    bookmarksnumbered,%
]{hyperref}

\newcommand{\white}{\color{white}\bf}

\begin{document}

\title{Microstructure and mechanical properties of\\constrained shape-memory alloy nanograins and nanowires}

\author{Mathieu Bouville}
	\email{m-bouville@imre.a-star.edu.sg}
	\affiliation{Institute of Materials Research and Engineering, Singapore 117602}
	\affiliation{Institute of High Performance Computing, Singapore 117528}

\author{Rajeev Ahluwalia}%
	\email{a-rajeev@imre.a-star.edu.sg}
	\affiliation{Institute of Materials Research and Engineering, Singapore 117602}
	\affiliation{Institute of High Performance Computing, Singapore 117528}

\begin{abstract}
We use the phase-field method to study the martensitic transformation at the nanoscale. For nanosystems such as nanowires and nanograins embedded in a stiff matrix, the geometric constraints and boundary conditions have an impact on martensite formation, leading to new microstructures ---such as dots aligned on a square lattice with axes along $\langle 01 \rangle$--- or preventing martensite formation altogether. We also perform tension tests on the nanowires. The stress--strain curves are very different from bulk results. Moreover, they are weakly affected by microstructures~--- the mechanical response of nanowires with different microstructures may be similar, while nanowires with the same microstructure may have a different mechanical behavior. We also observe that at the transition temperature, or slightly below it, the narrowest wires behave pseudoelastically whereas wider wires are in the memory-shape regime. Moreover the yield stress does not change monotonically with width: it has a minimum value at intermediate width.
\end{abstract}
\keywords{nanomaterials, phase-field simulations, mechanical testing, shape-memory, nanowires}
\pacs{
61.46.-w, 
62.20.fg, 
62.23.Hj, 
62.25.-g, 
64.70.Nd, 
81.30.Kf, 
81.70.Bt 
}
\maketitle

\section{Introduction}
Martensitic transformations are displacive (diffusionless) phase transformations from a high temperature high-symmetry austenite phase (usually cubic), to a low temperature low-symmetry phase known as the martensite phase (typically tetragonal, orthorhombic, or trigonal). Thus the transformation is accompanied by strain. Minimization of elastic energy in the martensite phase results in formation of a complex microstructure of the crystallographic variants.

The phase transformation and the complex microstructure are responsible for unusual mechanical properties, which make materials undergoing martensitic transformations useful for many technological applications.\cite{Otsuka-book, Bhattacharya-book} Martensite can exhibit the shape-memory effect, i.e.\ the existence of a residual strain upon unloading that can be recovered upon heating. It can also have a pseudoelastic behavior~--- a macroscopic deformation which is completely recovered when the load is removed. In the shape-memory regime, stress--strain curves are characterized by a residual strain when the stress goes down to zero after unloading, whereas with pseudoelasticity there is no residual strain and the system reverts to its initial state.

While martensitic transformations and microstructures in bulk are well understood, the behavior of these transformations at the nanoscale still requires investigation. For thin films,~\cite{Ishida-93, Krulevitch-MEMS-96, Bhattacharya-99, Walker-90}
nanowires,~\cite{Park-05, Liang-05} and
nanocrystals~\cite{Waitz-acta_mater-04, Wang-scripta_Mat-06}
the geometric constraints will have an impact on martensite formation. Recently, \citet{Frick-acta_mat-06} experimentally studied the stress--strain behavior of nanopillars of nickel titanium shape-memory alloys (SMA), using nanoindentation tests. They found that stress-induced martensitic transformation initiates at relatively low stresses. \Citet{Wang-scripta_Mat-06} studied the phase transformation behavior of grains of nickel titanium embedded in an amorphous matrix at different grain sizes. They found that the nature of the phase transformation changes, indicating that geometry and boundary conditions play a key role. Similarly, some applications of martensites make use of a  SMA element which is embedded in a matrix of a non-transforming material such as a polymer.~\cite{Bollas-acta_mater-07} 
Since these nanosystems are embedded in a matrix, there will be geometrical constraints on the phase transformation and microstructures: the martensite variants should arrange so that there is no macroscopic shape change. Geometry will also affect mechanical behavior. This leads us to the particular question of the effect of the size of the nanograins and nanowires~--- the smallest systems are expected to behave very differently from the bulk. An understanding of the size effects in such systems may provide insights for further research into nanoscale devices that use shape-memory materials. 

The effect of clamping~\cite{Jacobs-PRB-95,Jacobs-PRB-00} and non-transforming layers~\cite{Artemev-05, Seol-02} has been studied theoretically. However, the influence of geometry and size has not been systematically studied. We use the phase-field method to study the effect of size, shape, and transformation strains on the formation, microstructure, and mechanical properties of martensite in the case of nanowires and nanograins embedded in a non-transforming matrix.

\begin{figure}
\centering
\setlength{\unitlength}{1cm}
\begin{picture}(8.5,6.3)(.1,0)
\shortstack[c]{
\subfigure{
    \label{wire-200_0}
    \includegraphics[width=8.5cm]{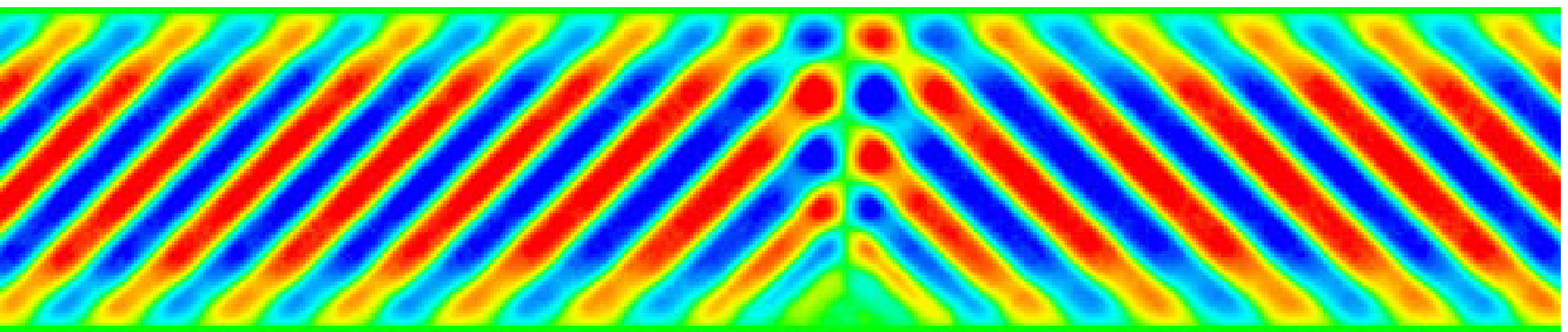}
	\put(-0.65, 1.3){\white(a)}
}\vspace{-3mm}\\
\subfigure{
    \label{wire-90_0}
    \includegraphics[width=8.5cm]{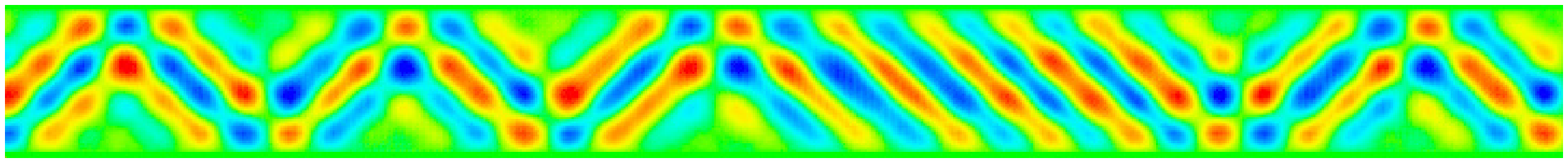}
	\put(-0.65, 0.35){\bf(b)}
}\vspace{-3mm}\\
\subfigure{
    \label{wire-86_2}
    \includegraphics[width=8.5cm]{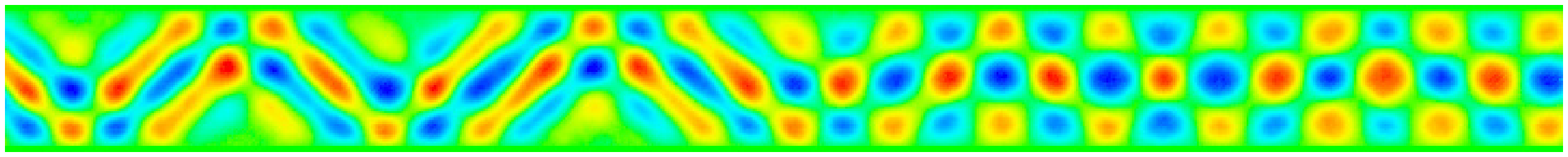}
	\put(-0.65, 0.35){\bf(c)}
}\vspace{-3mm}\\
\subfigure{
    \label{wire-72_0}
    \includegraphics[width=8.5cm]{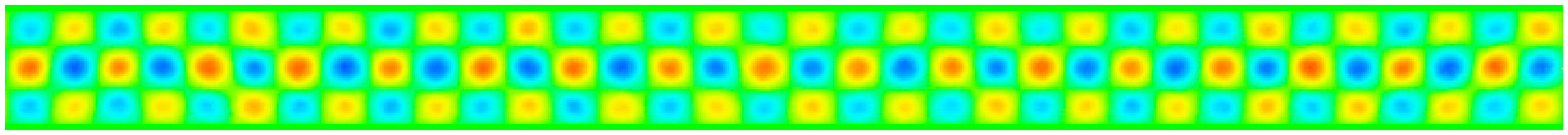}
	\put(-0.65, 0.25){\bf(d)}
}\vspace{-3mm}\\
\subfigure{
    \label{wire-71_0}
    \includegraphics[width=8.5cm]{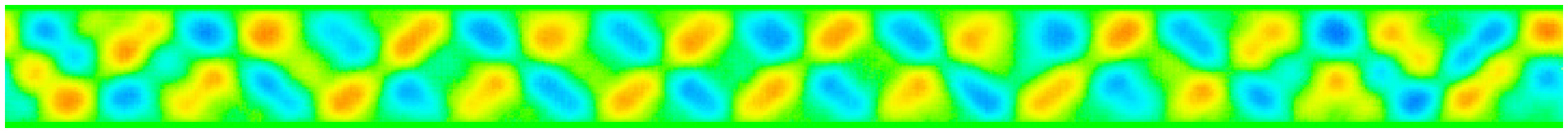}
	\put(-0.65, 0.25){\bf(e)}
}\vspace{-3mm}\\
\subfigure{
    \label{wire-68_0}
    \includegraphics[width=8.5cm]{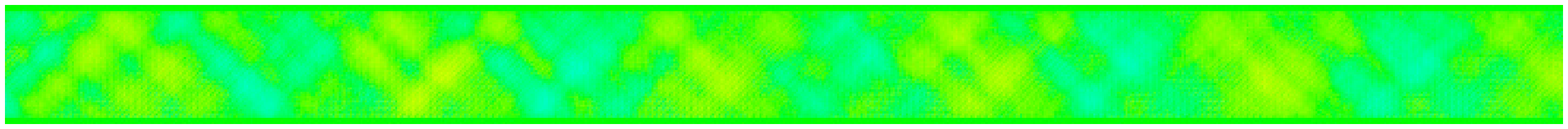}
	\put(-0.65, 0.25){\bf(f)}
}}
\end{picture}
	\caption{\label{nanowires_0}(Color online) Microstructure of nanowires for several values of the width $w$. \mbox{$x_{12}= 0$} (except in c, where \mbox{$x_{12}= 2$}).
(a)~$w =200$~nm,		(b)~$w = 90$~nm,
(c)~$w = 86$~nm,		(d)~$w = 72$~nm,
(e)~$w = 71$~nm, and	(f)~$w = 68$~nm. 
Green: austenite; red and blue: martensite.}
\end{figure}

\section{Phase-field model}
The free energy of our two-dimensional system is based on the usual non-linear elastic free energy density for a square-\linebreak[3]to-\linebreak[3]rectangle martensitic transition:~\cite{Falk80, Onuki-99}
\begin{eqnarray}
	&G=\int \left\{ \dfrac{A_{22}}{2}\dfrac{T\!-T_m}{T_m}(e_2)^2 - \dfrac{A_{24}}{4}(e_2)^4 + \dfrac{A_{26}}{6}(e_2)^6 + \right. \nonumber\\
	&  \left. \dfrac{A_1}{2} \left[ e_1 - x_{12}\,(e_2)^2  \right]^2 + \dfrac{A_3}{2} (e_3)^2
+ \dfrac{k}{2} \|\nabla e_2\|^2 \right\} \mathrm{d}\mathbf{r}.
	\label{eq-G}
\end{eqnarray}
\noindent Here $T$ is the temperature, $T_m$ is the austenite--martensite transition temperature, $e_1$ is the hydrostatic strain, $e_2$ is the deviatoric strain, and $e_3$ is the shear strain:
\begin{subequations}
\begin{align}
	e_1 &= (\varepsilon_{xx}+\varepsilon_{yy})/\sqrt{2},\\
	e_2 &= (\varepsilon_{xx}-\varepsilon_{yy})/\sqrt{2},\\
	e_3 &= \varepsilon_{xy}.
\end{align}
\end{subequations}
\noindent The $\{\varepsilon_{ij}\}$ are the linearized strain tensor components. In the case of a square lattice,  $\varepsilon_{xx}=\varepsilon_{yy}$ so that $e_2=0$. This corresponds to austenite. In the case of martensite, $\varepsilon_{xx} \ne \varepsilon_{yy}$ and $e_2\ne0$. The deviatoric strain $e_2$ is thus used to track austenite and martensite.

To lowest order in $e_2$, the energy is quadratic with an extremum at $e_2=0$. For temperatures above $T_m$, the first term in Eq.~(\ref{eq-G}) is positive, so that $e_2=0$ is a minimum, i.e.\ austenite is (meta)stable. On the other hand, if $T<T_m$ the first term in Eq.~(\ref{eq-G}) is negative and $e_2=0$ is a maximum: austenite is unstable.

The evolution of the displacements is described by~\cite{landau-lifschitz, Ahluwalia-acta_mater-06}
\begin{equation}
	\rho \,\frac{\partial^2\, u_i(\mathbf{r}, t)}{\partial\, t^2} = 
\sum_j\frac{\partial\, \sigma_{ij}(\mathbf{r}, t)}{\partial\, r_j} +
\eta \,\mathbf{\nabla}^2 v_i(\mathbf{r}, t),
\label{du_dt}
\end{equation}

\noindent where $\rho$ is a density, $\mathbf{v}$ is the time derivative of the displacements $\mathbf{u}$, and the stresses are given by
\begin{equation}
	\sigma_{ij}(\mathbf{r}, t) = \frac{\delta\, G}{\delta\, \varepsilon_{ij}(\mathbf{r}, t)},
\end{equation}

\noindent with $\delta$ the functional derivative. The second term on the right-hand side in Eq.~(\ref{du_dt}) is a viscous damping term; it is a simplification of the more general damping of Ref.~\onlinecite{Ahluwalia-acta_mater-06}. 

We choose values for the parameters corresponding to FePd:\cite{Kartha95} $A_1 = 140$~GPa, $A_3 = 280$~GPa, $A_{22} = 212$~GPa, $A_{24} = 17 \times 10^3$~GPa, $A_{26} = 30 \times 10^6$~GPa, and $T_m=265$~K. We set $k$ to 200 and $\eta$ to 0.01.
The coupling constant $x_{12}$ is varied in different cases to understand the effect of volume changes, as in Refs.~\onlinecite{Bouville-PRL-06} and~\onlinecite{Bouville-PRB-07}. 

In each two-dimensional simulation, austenite is quenched to $T=250~\text{K}<T_m$ and the system is allowed to transform and equilibrate. The stiff matrix in which the nanosystem is embedded is simulated by fixing the displacements to zero at the boundary (for nanowires there are periodic boundaries along the axis of the wire).

\begin{figure}
\centering
\setlength{\unitlength}{1cm}
\begin{picture}(8.5,3.4)(.1,0)
\shortstack[c]{
\subfigure{
    \label{wire-90_10}
    \includegraphics[width=8.5cm]{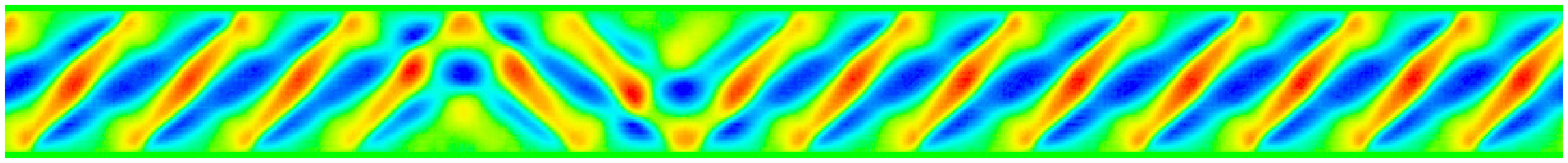}
	\put(-0.65, 0.35){\white(a)}
}\vspace{-3mm}\\
\subfigure{
    \label{wire-74_8}
    \includegraphics[width=8.5cm]{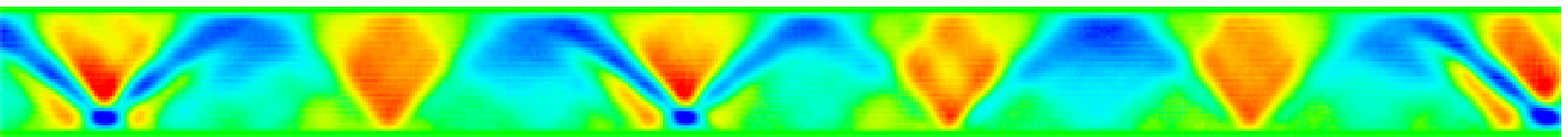}
	\put(-0.65, 0.25){\bf(b)}
}\vspace{-3mm}\\
\subfigure{
    \label{wire-72_9}
    \includegraphics[width=8.5cm]{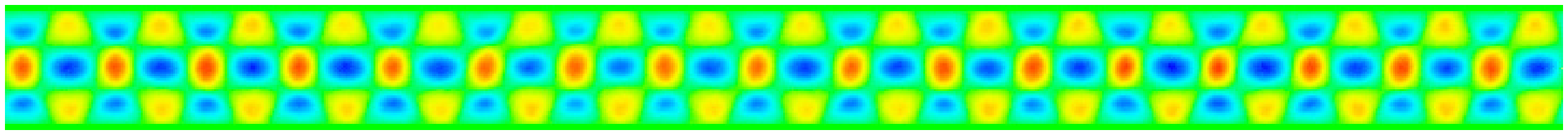}
	\put(-0.65, 0.25){\bf(c)}
}\vspace{-3mm}\\
\subfigure{
    \label{wire-70_10}
    \includegraphics[width=8.5cm]{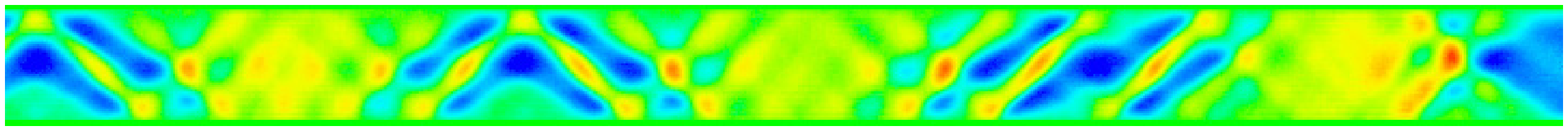}
	\put(-0.65, 0.25){\bf(d)}
}}
\end{picture}
	\caption{\label{nanowires_10}(Color online) Microstructure of nanowires for several values of the width $w$ and of the volume change $x_{12}$. 
(a)~$w = 90$~nm, \mbox{$x_{12}=10$}; 
(c)~$w = 74$~nm, \mbox{$x_{12}= 8$}; 
(b)~$w = 72$~nm, \mbox{$x_{12}= 9$}; and
(d)~$w = 70$~nm, \mbox{$x_{12}=10$}. 
Green: austenite; red and blue: martensite.}
\end{figure}

\section{Microstructures}

\subsection{Nanowires --- microstructure}
Figures~\ref{nanowires_0} and~\ref{nanowires_10} show the microstructure of nanowires (for low and high values of $x_{12}$ respectively). The wires are 2~$\mu$m long; figures show only one half of the wire, for clarity. Here and in all other figures, austenite is shown in green and the two martensite variants are in red and blue, respectively. The widest wires show twins similar to what is observed in bulk martensite, Fig.~\ref{wire-200_0}. One can see that the twins do not extend unhindered to the interface. As long as this zone that differs from the bulk is quite thinner than the wire the microstructure is that of the bulk. But in the case of narrower wires, finite size effects dominate.
For very narrow wires, the martensitic transformation is completely suppressed, Fig.~\ref{wire-68_0}. For intermediate widths, e.g.\ Fig.~\ref{wire-72_0}, the system exhibits a microstructure that is not observed in bulk systems:~\cite{Bouville-PRL-06, Bouville-PRB-07} dots aligned on a square lattice with axes along $\langle 01 \rangle$. For these widths, the transformation is not completely suppressed and thus we observe retained austenite along with the two martensite variants, especially in Fig.~\ref{wire-71_0}.

\begin{figure}
\centering
\setlength{\unitlength}{1cm}
\begin{picture}(8.5,18.4)(.1,0)
\shortstack[c]{
\subfigure{
    \label{diag-wire}
    \includegraphics[width=8.5cm]{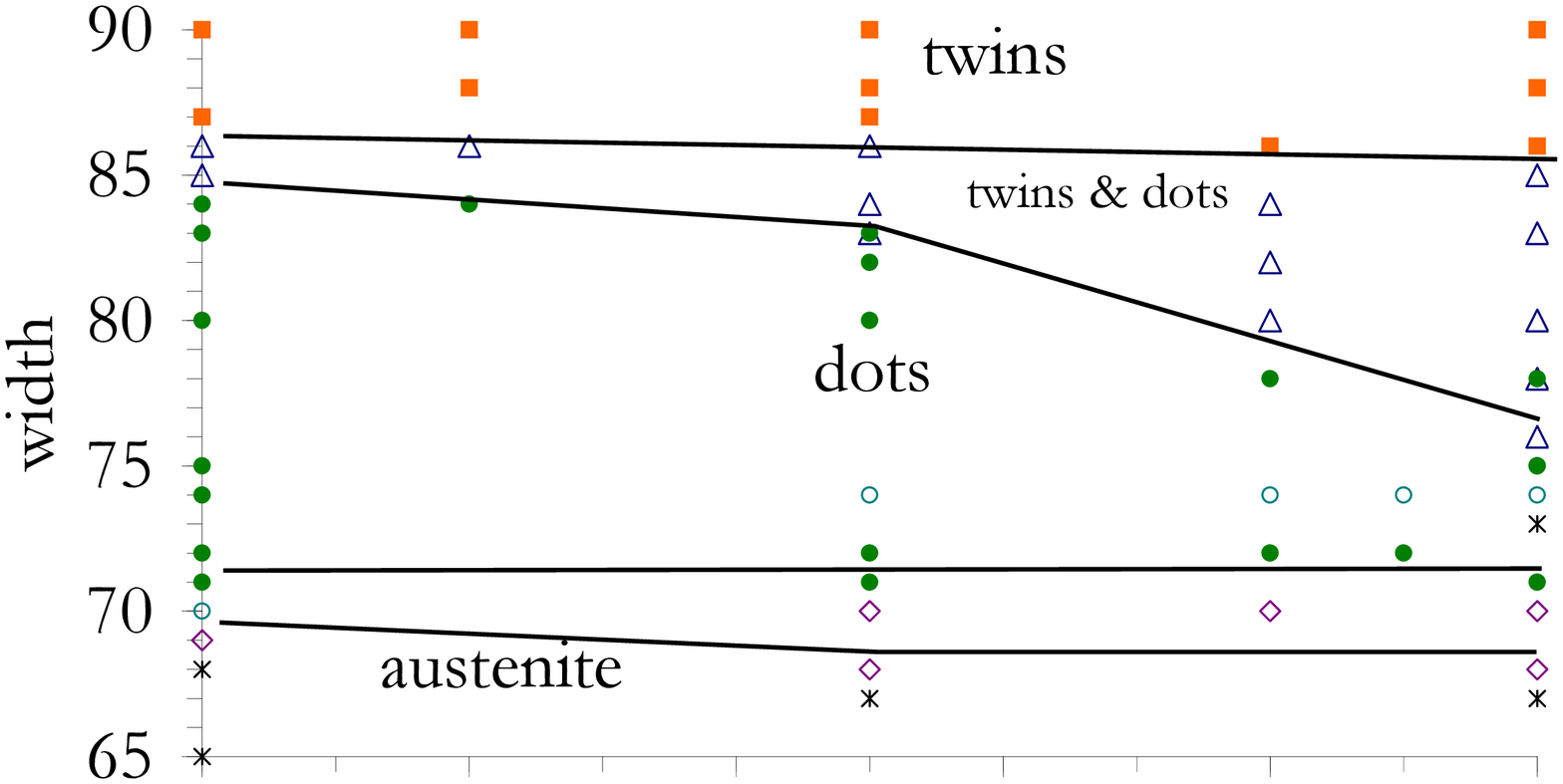}
	\put(-7.2, 3.8){\bf(a)}
}\vspace{-2mm}\\
\subfigure{
    \label{diag-square}
    \includegraphics[width=8.5cm]{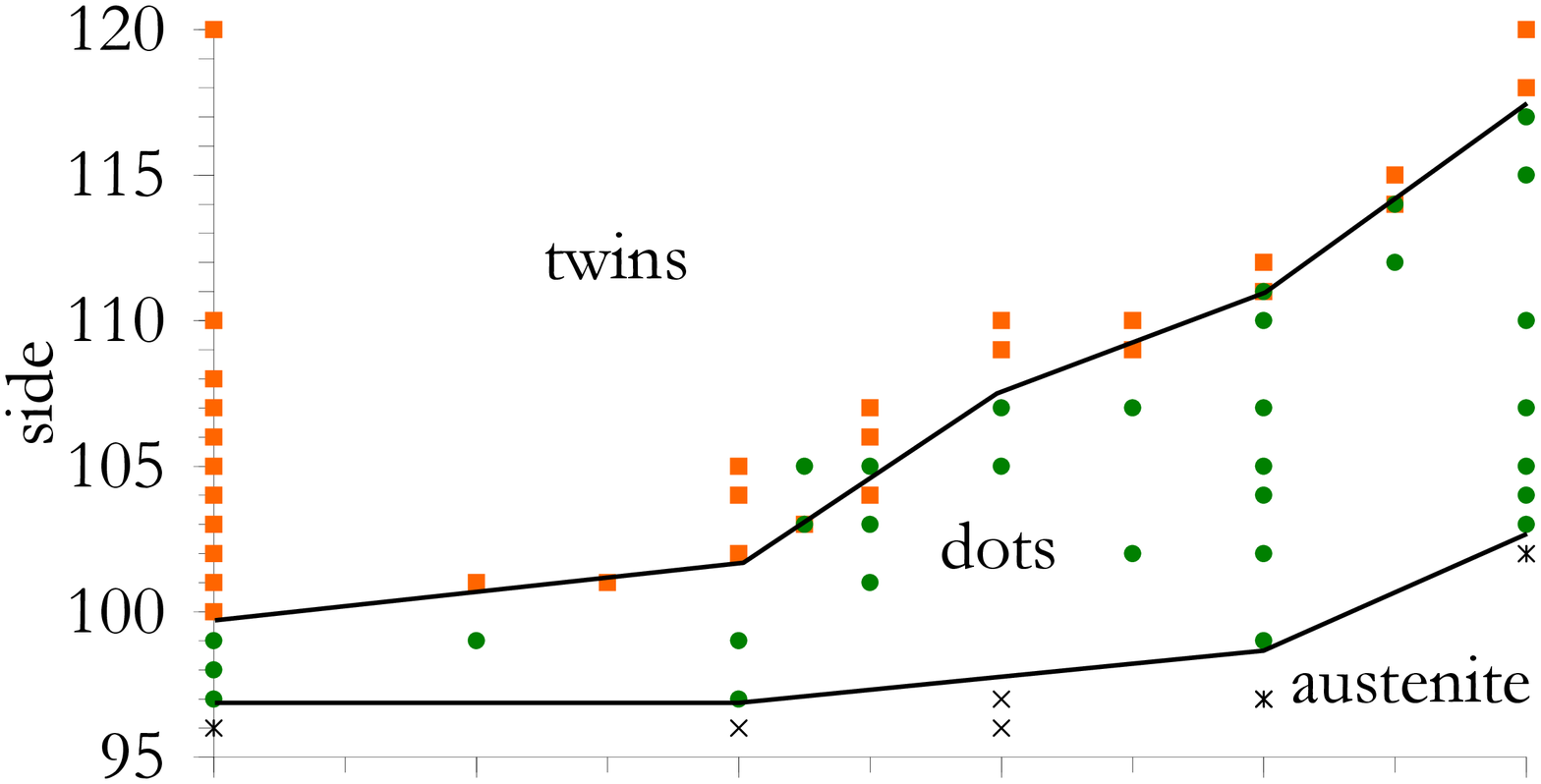}
	\put(-7.2, 3.8){\bf(b)}
}\vspace{-2mm}\\
\subfigure{
    \label{diag-tilted}
    \includegraphics[width=8.5cm]{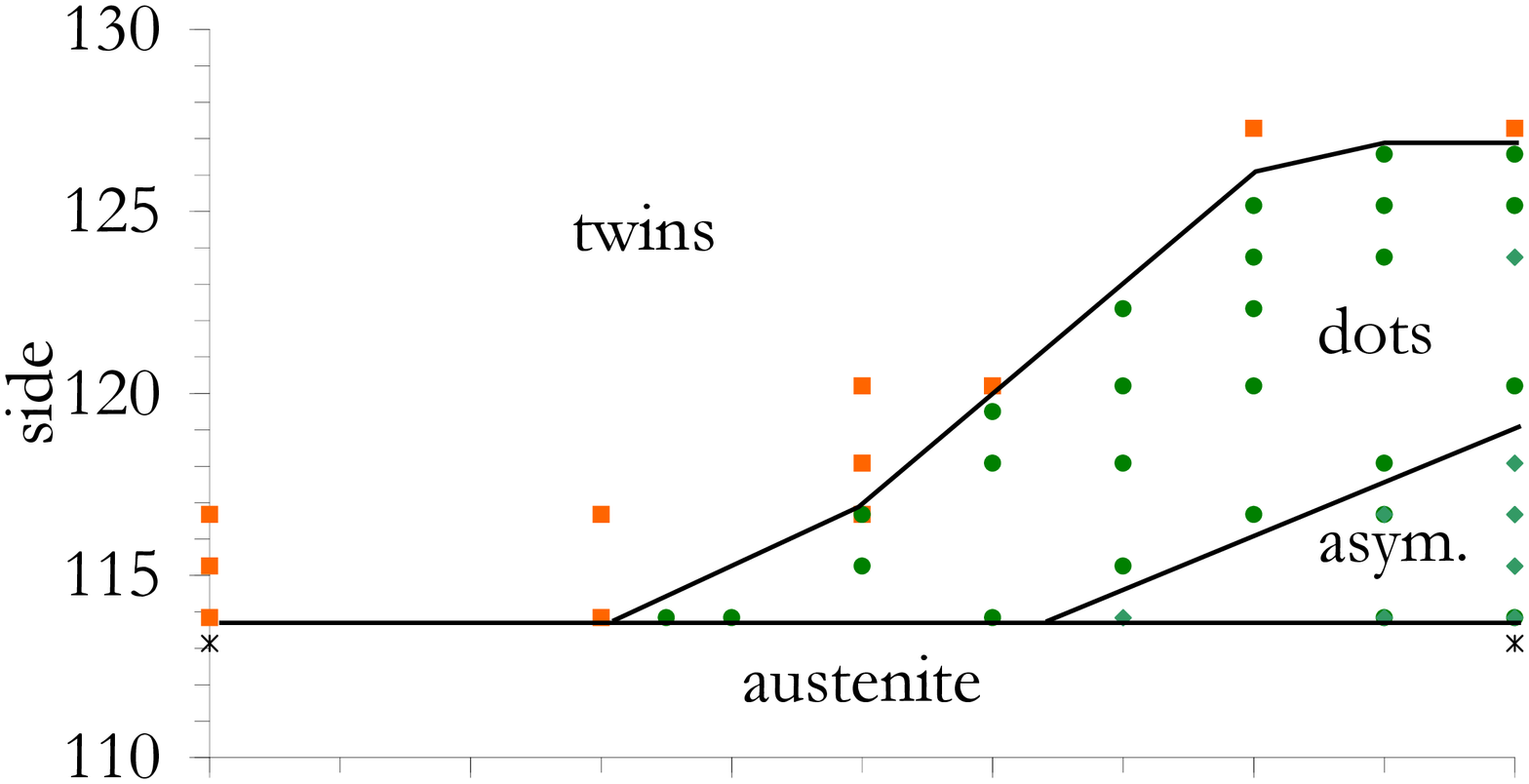}
	\put(-7.2, 3.8){\bf(c)}
}\vspace{-2mm}\\
\subfigure{
    \label{diag-circ}
    \includegraphics[width=8.5cm]{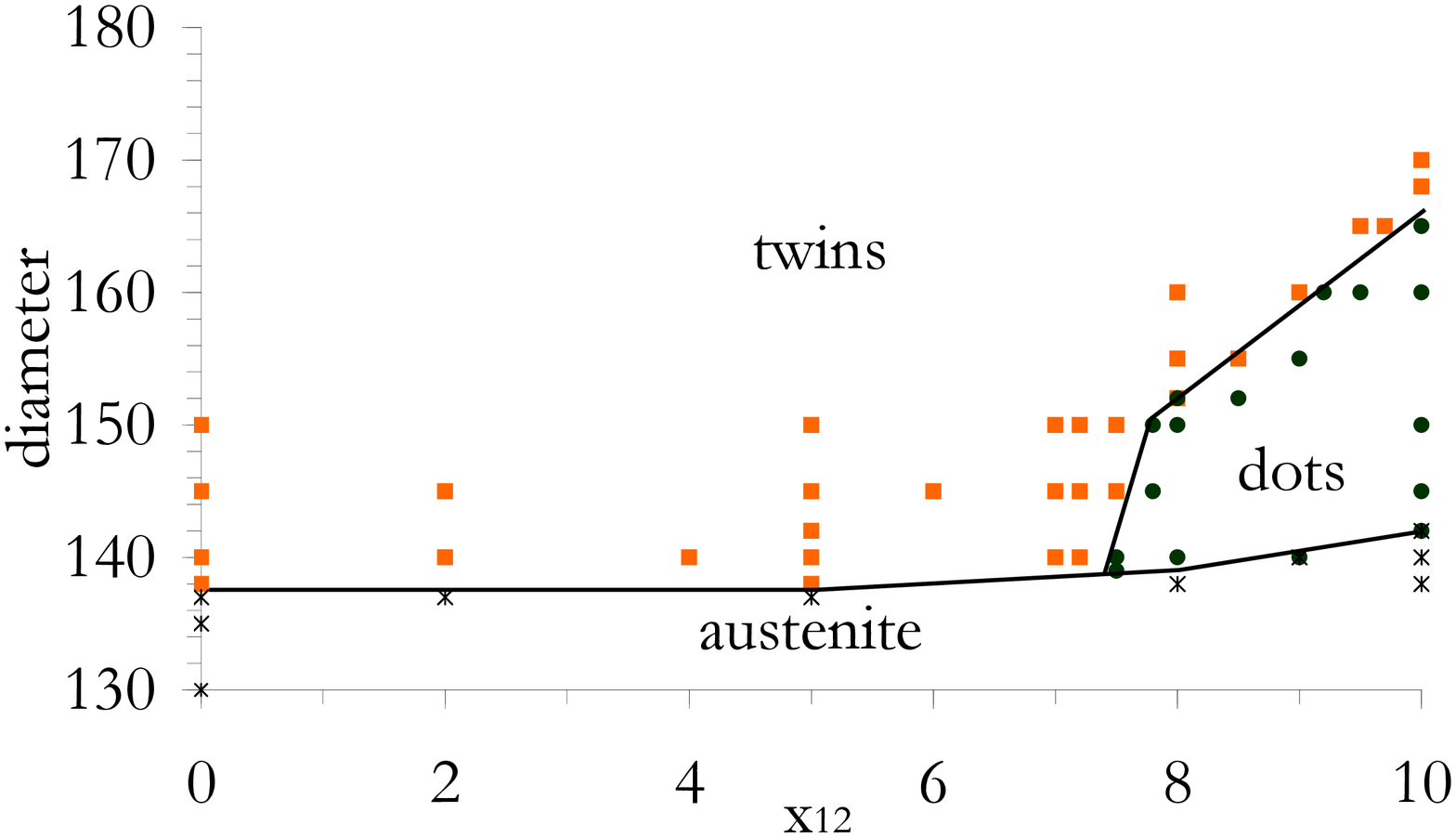}
	\put(-7.2, 4.3){\bf(d)}
}}
\end{picture}
	\caption{\label{diag}Microstructure as a function of volume change $x_{12}$ and of the system size (in nm). (a)~nanowire, (b)~square grain with faces along $\langle 01 \rangle$, (c)~square grain with faces along $\langle 11 \rangle$, (d)~circular grain.}
\end{figure}

In some materials systems, there is a significant volume change associated with the transformation (the volume of the martensite unit cell is different from that of austenite), so that the phase transformation generates hydrostatic stress, thus affecting the microstructure.~\cite{Bouville-PRL-06, Bouville-PRB-07} Figures~\ref{wire-90_10} and~\ref{wire-72_9} show that in the presence of a volume change, microstructures change somewhat: dots may become triangular and/or elongated and twins are no longer parallel. 

There is also a range of widths (especially at large $x_{12}$) exhibiting a mixture of twins and dots, Fig.~\ref{wire-86_2}. For intermediate widths, we see clustered variants, Figs.~\ref{wire-74_8} and~\ref{wire-70_10}. For $w=71$~nm, the system exhibits dots, but these do not reach the usual strain associated with martensite and there are only two of them across the wire rather than three, as shown in Fig.~\ref{wire-71_0}.
We obtain a `microstructure diagram' \mbox{---akin} to a phase diagram--- showing the microstructure (austenite, twins, dots) as a function of the width of the wire and of volume change, Fig.~\ref{diag-wire}. There are five possible microstructures for decreasing sizes: twins, twin plus dots, dots, clustered variants, and austenite. The volume change $x_{12}$ plays a minor role in determining microstructures.

\begin{figure}
\centering
\setlength{\unitlength}{1cm}
\begin{picture}(8.5,2.8)(.1,0)
\subfigure{
    \label{sq-120_0}
    \includegraphics[width=2.8cm]{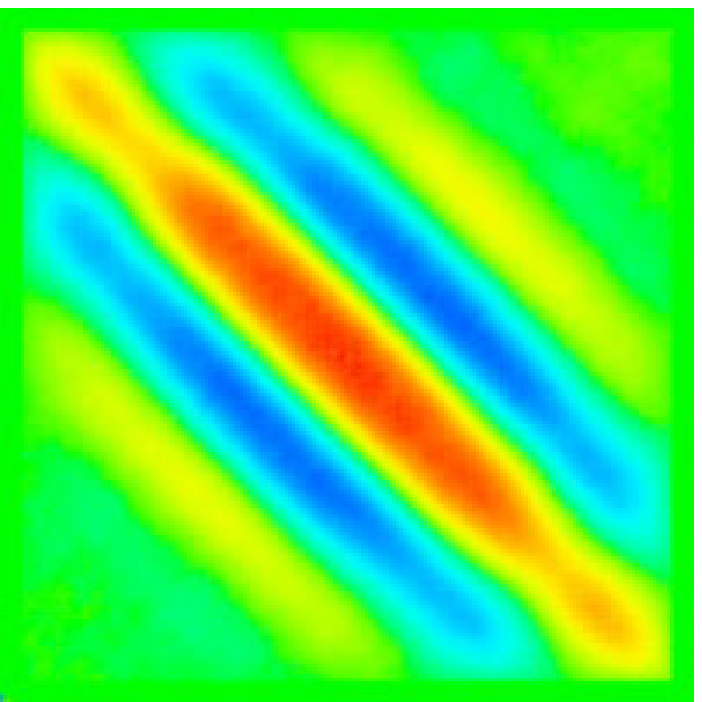}
	\put(-0.6, 2.3){\bf(a)}
}\subfigure{
    \label{sq-103_45}
    \includegraphics[width=2.8cm]{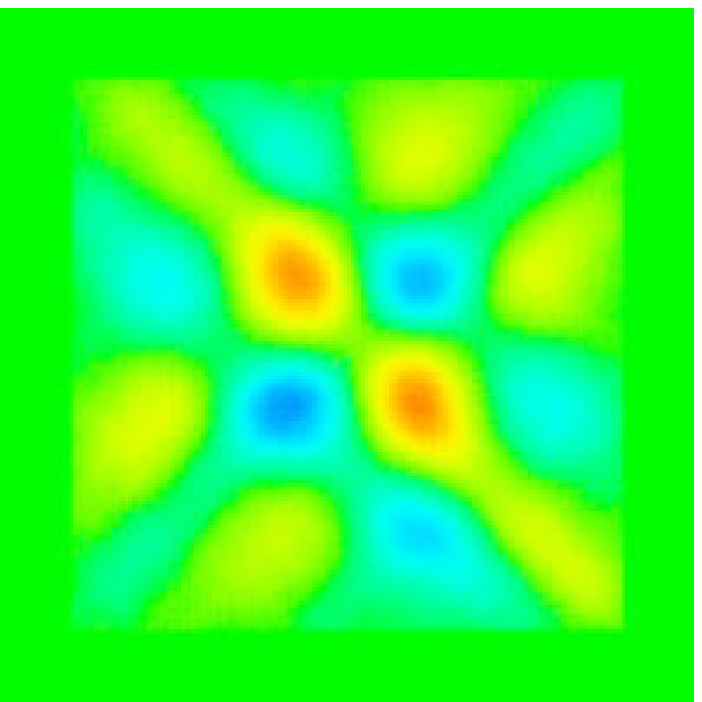}
	\put(-0.6, 2.3){\bf(b)}
}\subfigure{
    \label{sq-120_10}
    \includegraphics[width=2.8cm]{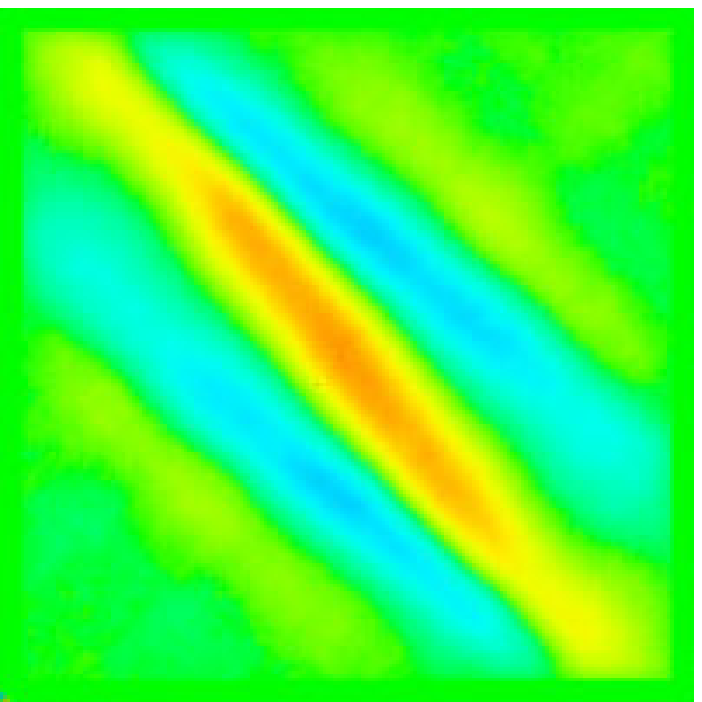}
	\put(-0.6, 2.3){\bf(c)}
}
\end{picture}
	\caption{\label{square}(Color online) Microstructures for a square grain with faces along $\langle 01 \rangle$. (a)~$\text{side}=120$~nm, $x_{12}=0$; (b)~$\text{side}=103$~nm, $x_{12}=4.5$; and (c)~$\text{side}=120$~nm, $x_{12}=10$.}
\end{figure}

\subsection{Nano grains}
Similar work is carried out for nanoscale grains embedded in a non-transforming matrix. We study their microstructure as a function of shape, size and orientation, as well as volume change. We consider the cases of circular grains and of square grains (with sides oriented either along the $\langle 01 \rangle$ direction of the matrix or along the $\langle 11 \rangle$ direction).

Figure~\ref{square} shows microstructures and Fig.~\ref{diag-square} the microstructure diagram for a square grain with faces along $\langle 01 \rangle$ (note that Fig.~\ref{sq-120_0} is similar to the results of \citet{Jacobs-PRB-95}). The trend is similar to nanowires with twins, dots, and austenite succeeding as the grain shrinks. A noteworthy difference between Figs.~\ref{diag-square} and~\ref{diag-wire} is that the dependence on $x_{12}$ is very weak in the case of nanowire but noticeable for square grains. Also, no coexistence of twins and dots can be observed because grains are to small to have several domains (they are small along both directions, whereas nanowires are nano only along one direction).

\begin{figure}
\centering
\setlength{\unitlength}{1cm}
\begin{picture}(8.5,2.8)(.1,0)
\subfigure{
    \label{tilted-175_0}
    \includegraphics[width=2.8cm]{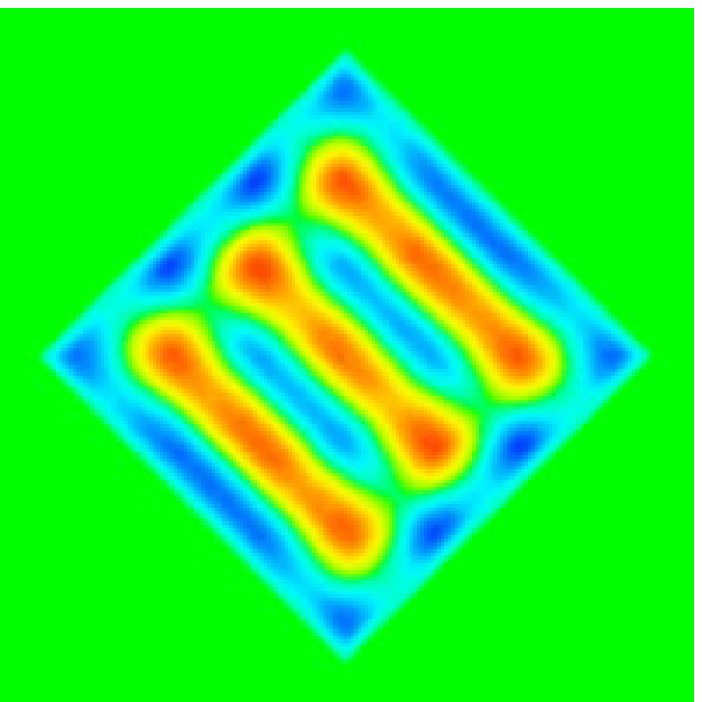}
	\put(-0.6, 2.3){\bf(a)}
}\subfigure{
    \label{tilted-189_10}
    \includegraphics[width=2.8cm]{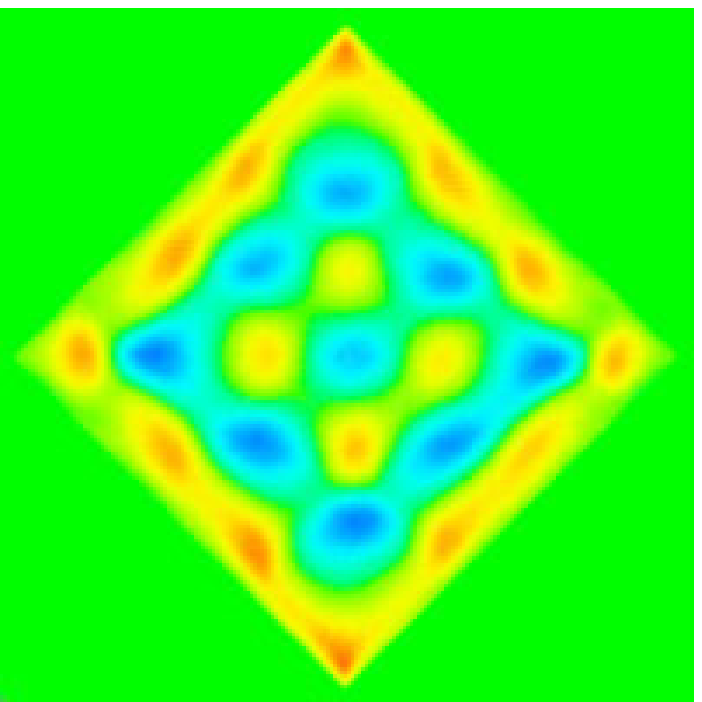}
	\put(-0.6, 2.3){\bf(b)}
}\subfigure{
    \label{tilted-177_10}
    \includegraphics[width=2.8cm]{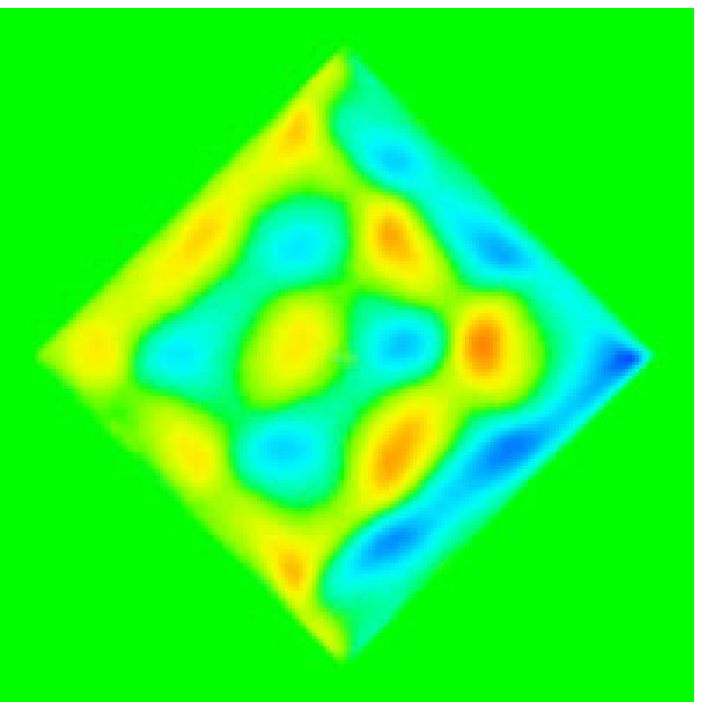}
	\put(-0.6, 2.3){\bf(c)}
}
\end{picture}
	\caption{\label{tilted}(Color online) Microstructures for a square grain with faces along $\langle 11 \rangle$. (a)~$\text{side}= 116.7$~nm, $x_{12}=0$; (b)~$\text{side}= 126.6$~nm, $x_{12}=10$; (c)~$\text{side} = 118.1$~nm, $x_{12}=10$.}
\end{figure}

This dependence on $x_{12}$ is even stronger in the case of a square grain with faces along $\langle 11 \rangle$, Fig.~\ref{diag-tilted}: there can be no dot microstructure if $x_{12}=0$. Moreover, there exists an asymmetric dot microstructure for high $x_{12}$ and intermediate size, as shown in Fig.~\ref{tilted-177_10}. One can also notice that for square grains with faces along $\langle 11 \rangle$ the transition occurs for larger systems than in the case of square grains with faces along $\langle 01 \rangle$.

For circular grains, Fig.~\ref{diag-circ}, the dependence on $x_{12}$ is even stronger, with no dots for $x_{12}<7$. As in other cases, the martensite--austenite transition depends weakly on the volume change $x_{12}$.

\section{Mechanical testing}
In the previous section we have shown that nanowires and nanograins exhibit microstructures that are not found in the bulk. An obvious question is whether this will affect the mechanical properties of nanoscale systems. In this section and the next, we will perform tension tests on nanowires to determine how the mechanical response of nanowires differs from that of the bulk.

A wire of length 2~$\mu$m is annealed until the microstructure no longer evolves. It is then loaded at a constant strain rate along its axis (in tension) from time 0 to time 70 (arbitrary units), then unloaded. At every point the strain $\varepsilon_{xx}$ is given by 
\begin{equation*}
	\varepsilon_{xx} = \frac{\partial u_x}{\partial x} + \dot{\varepsilon}\,t.
\end{equation*}

In this section, 
the strain rate will be 0.05\% per time unit and there will be no volume change ($x_{12}=0$). In the next section, these two parameters will be varied to study their impact on microstructures and mechanical properties. Except in Sec.~\ref{sec-stress_max} the temperature will be $T=250$~K (which is lower that the transition temperature, $T_m=265$~K).

\subsection{Difference between nanowire and bulk}
In the simulations, the nanowire is embedded in a stiff matrix so that $\varepsilon_{yy}=0$. Consequently, $e_1 = e_2 = \varepsilon_{xx}/\sqrt{2}$. Assuming that $e_3$ is negligible and using that $x_{12}=0$, the energy becomes
\begin{equation*}
	g = \left(\!\dfrac{A_{22}}{4}\dfrac{T\!-T_m}{T_m} + \dfrac{A_1}{4} \!\right) (\varepsilon_{xx})^2 - \dfrac{A_{24}}{16}(\varepsilon_{xx})^4 +  
	  \dfrac{A_{26}}{48}(\varepsilon_{xx})^6.
\end{equation*}
\noindent Since $\sigma_{xx} = {\mathrm{d} g}/{\mathrm{d} \varepsilon_{xx}}$, we have
\begin{equation}
	 \sigma_{xx} =
	\left(\!\dfrac{A_{22}}{2}\dfrac{T\!-T_m}{T_m} + \dfrac{A_1}{2}\!\right)\varepsilon_{xx} 
	 - \dfrac{A_{24}}{4}(\varepsilon_{xx})^3 +  \dfrac{A_{26}}{8}(\varepsilon_{xx})^5.
	\label{eq-stress}
\end{equation}

\begin{figure}
	\centering
    \includegraphics[width=8.5cm]{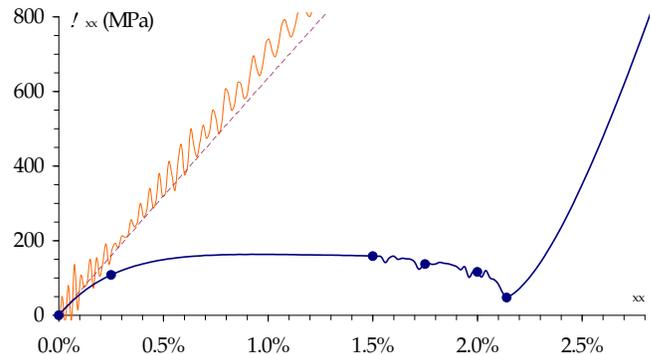}
	\caption{\label{theoretical}Stress--strain curves: analytical (dashed line), bulk (solid line), and nanowire of width $w=1000$~nm (thick line). (The dots correspond to the microstructures shown in Fig.~\ref{mech-1000_0}.)}
\end{figure}
 
We also performed a bulk simulation (with periodic boundary conditions) of the loading process, starting from a twinned martensitic state. Figure~\ref{theoretical} shows that the stress--strain curve for the bulk system is very close to the analytical result of Eq.(\ref{eq-stress}). The bulk system eventually reaches a single variant; on unloading it follows the same stress--strain curve, so that there is no residual strain.

\begin{figure}
\centering
\setlength{\unitlength}{1cm}
\begin{picture}(8.5,15.5)(.1,0)
\shortstack[c]{
\subfigure{
    \label{mech-1000_0-t0000}
    \includegraphics[width=8.5cm]{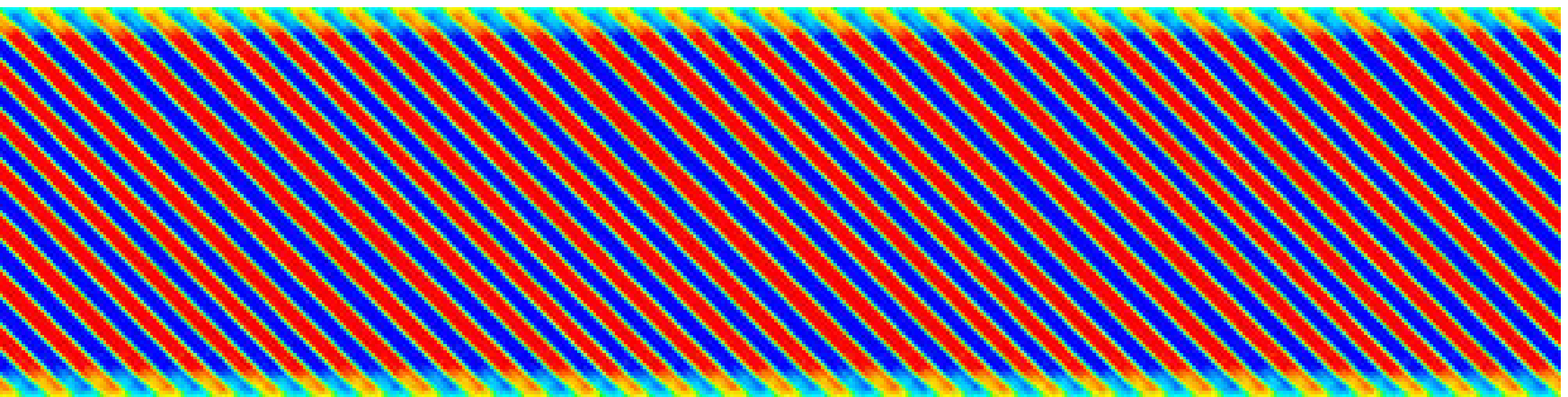}
	\put(-0.6, 1.7){\white(a)}
}\vspace{-3.5mm}\\
\subfigure{
    \label{mech-1000_0-t0005}
    \includegraphics[width=8.5cm]{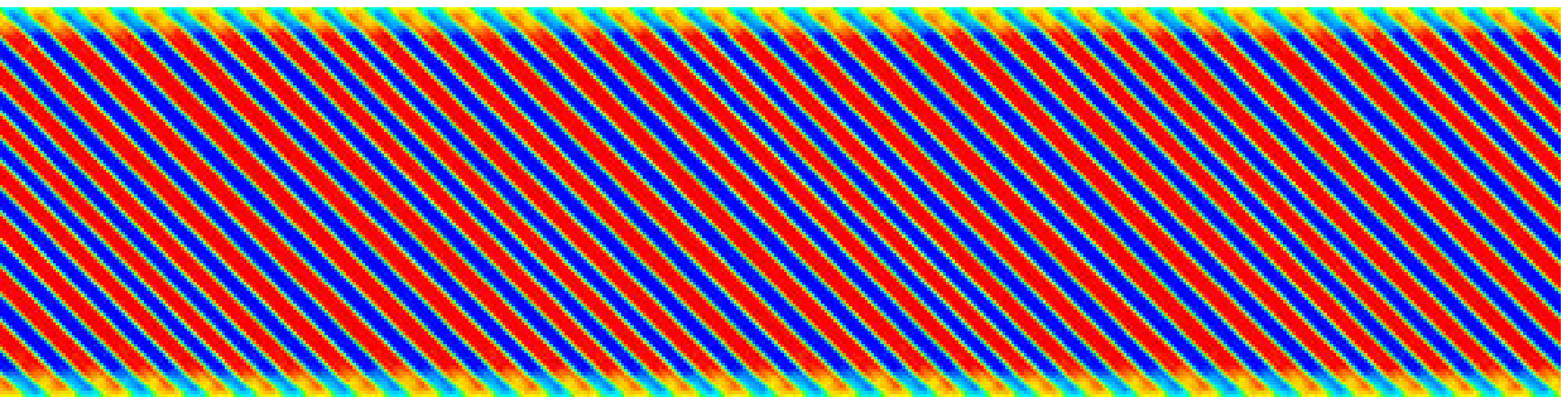}
	\put(-0.6, 1.7){\white(b)}
}\vspace{-3.5mm}\\
\subfigure{
    \label{mech-1000_0-t0030}
    \includegraphics[width=8.5cm]{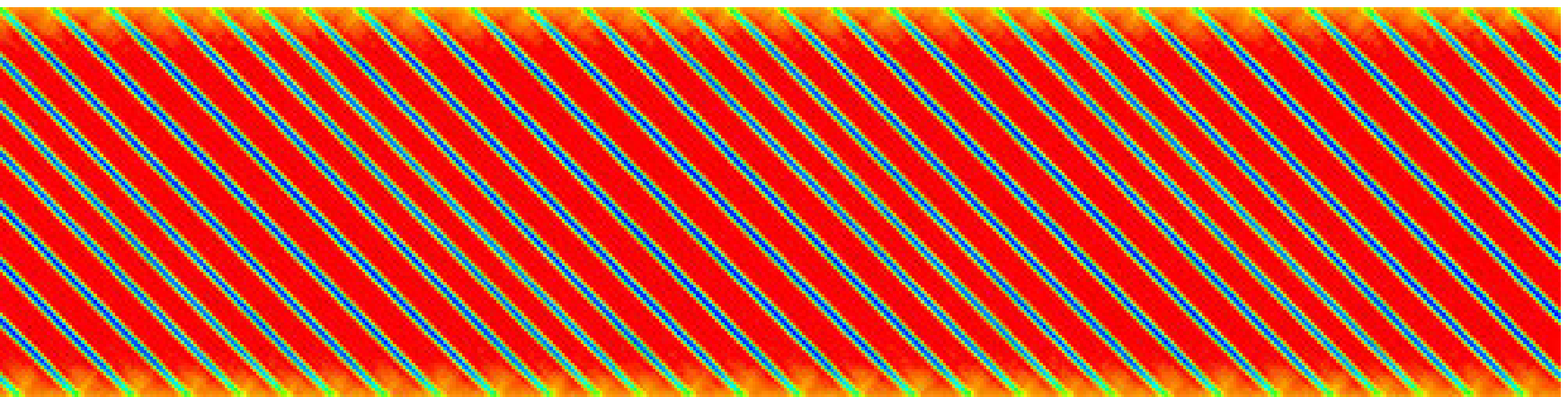}
	\put(-0.6, 1.7){\bf(c)}
}\vspace{-3.5mm}\\
\subfigure{
    \label{mech-1000_0-t0035}
    \includegraphics[width=8.5cm]{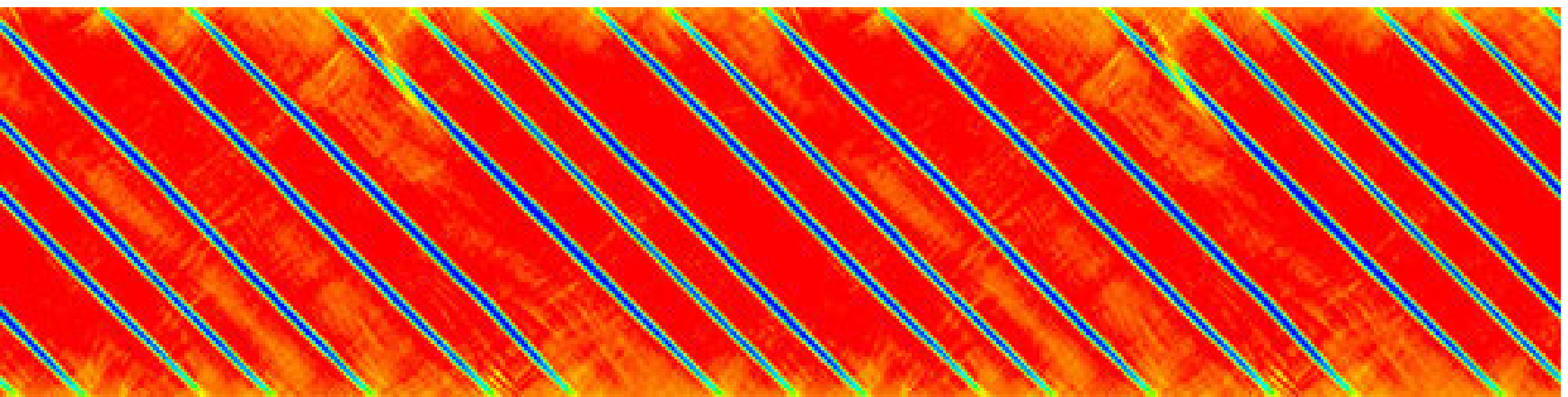}
	\put(-0.6, 1.7){\bf(d)}
}\vspace{-3.5mm}\\
\subfigure{
    \label{mech-1000_0-t0040}
    \includegraphics[width=8.5cm]{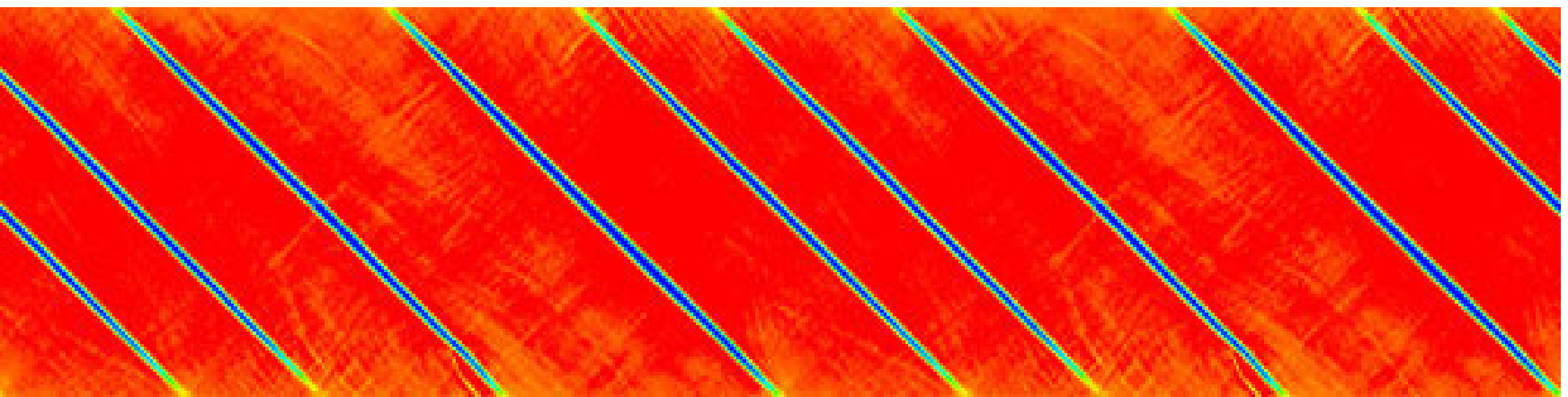}
	\put(-0.6, 1.7){\bf(e)}
}\vspace{-3.5mm}\\
\subfigure{
    \label{mech-1000_0-t0043}
    \includegraphics[width=8.5cm]{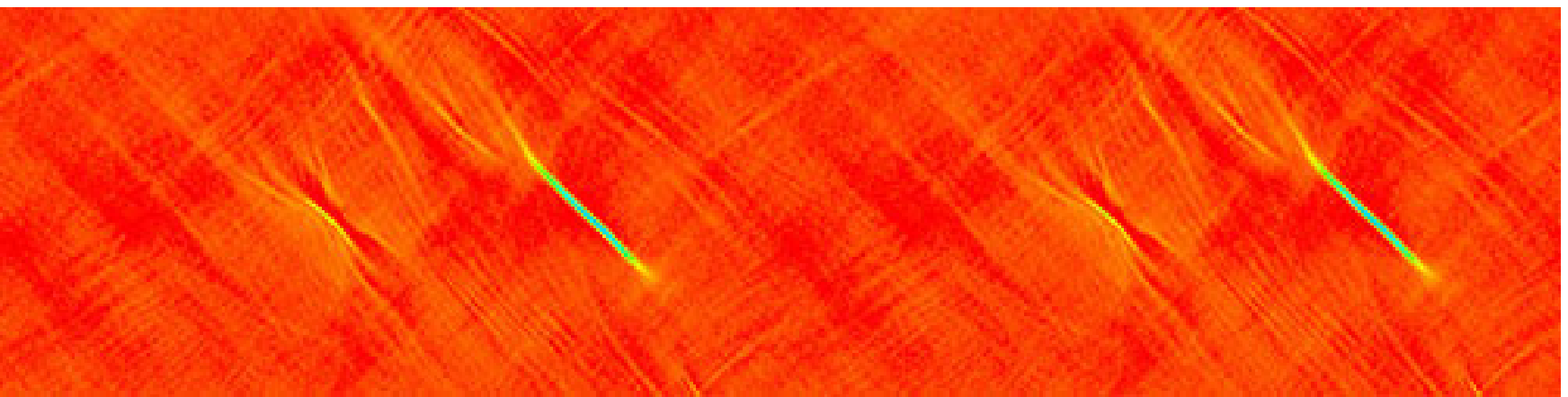}
	\put(-0.6, 1.7){\bf(f)}
}\vspace{-3.5mm}\\
\subfigure{
    \label{mech-1000_0-t0140}
    \includegraphics[width=8.5cm]{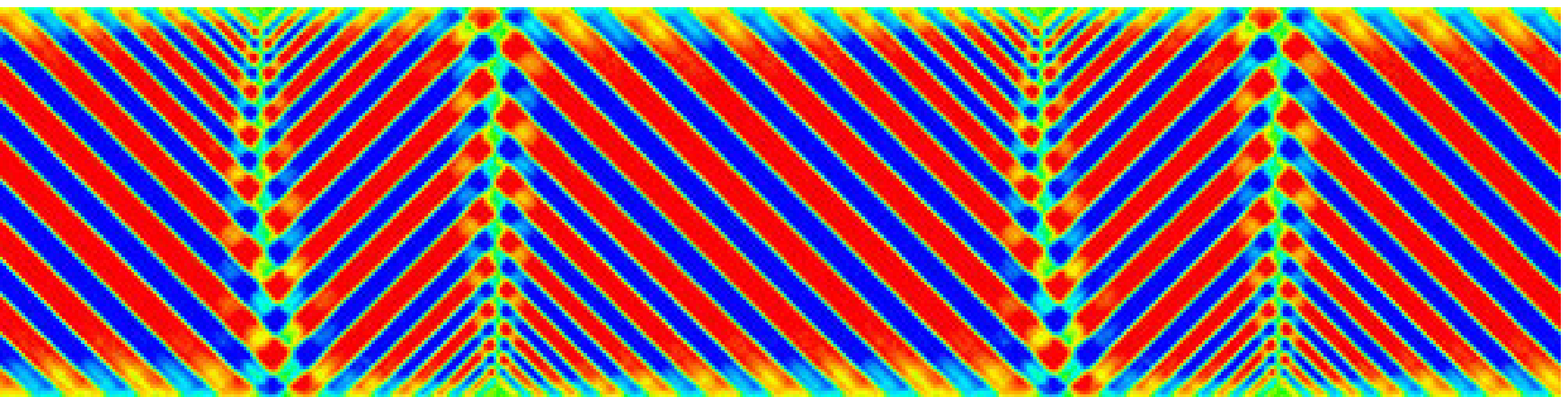}
	\put(-0.6, 1.7){\white(g)}
}}
\end{picture}
\caption{\label{mech-1000_0}(Color online) Microstructure of a nanowire of width $w=1000$~nm. 
(a)~$t=0$,  $\varepsilon = 0\%$; 
(b)~$t=5$, $\varepsilon = 0.25\%$; 
(c)~$t=30$, $\varepsilon = 1.5\%$; 
(d)~$t=35$, $\varepsilon = 1.75\%$; 
(e)~$t=40$, $\varepsilon = 2\%$; 
(f)~$t=43$, $\varepsilon = 2.15\%$; and
(g)~$t=140$, $\varepsilon = 0\%$. 
(For each time, the simulated system is shown twice over.)}
\end{figure}

Figure~\ref{theoretical} shows that the stress--strain curve for a relatively wide wire is very different from both the analytical solution and the bulk simulation. Notice that there is a plateau in the stress--strain curve. The microstructural evolution of this nanowire is shown in Fig.~\ref{mech-1000_0}. There exist two martensite variants, with positive and negative values of $e_2$ (corresponding to rectangular unit cells elongated along the x-axis and the y-axis respectively). When $\varepsilon_{xx}$ increases due to the loading, $e_2 = (\varepsilon_{xx}-\varepsilon_{yy})/\sqrt{2}$ also increases. The variant corresponding to \mbox{$e_2>0$} (in red in Fig.~\ref{mech-1000_0}) is thus more and more energetically favored while the variant with \mbox{$e_2<0$} (blue) becomes higher and higher in energy. The former thus grows and the latter shrinks, so that eventually a single variant remains.

In the linear part of the stress--strain curve for this nanowire, there is no change of microstructure: compare Figs.~\ref{mech-1000_0}(a) and~\ref{mech-1000_0}(b); this part corresponds to the elastic response of the martensite. The plateau corresponds to the growth of the favored variant (red) over the unfavored one (blue) at constant twin number, from Fig.~\ref{mech-1000_0}(b) to Fig.~\ref{mech-1000_0}(c). Then some of the unfavored twins start disappearing and the system loses its periodicity, Figs.~\ref{mech-1000_0}(d--f), which results in oscillations in the stress--strain curve. Figure~\ref{mech-1000_0}(f) corresponds to the point when the unfavored variant is about to disappear completely. After that, the stress--strain curve merely shows the elastic response of a single martensite variant.

\begin{figure}
	\centering
	\includegraphics[width=8.6cm]{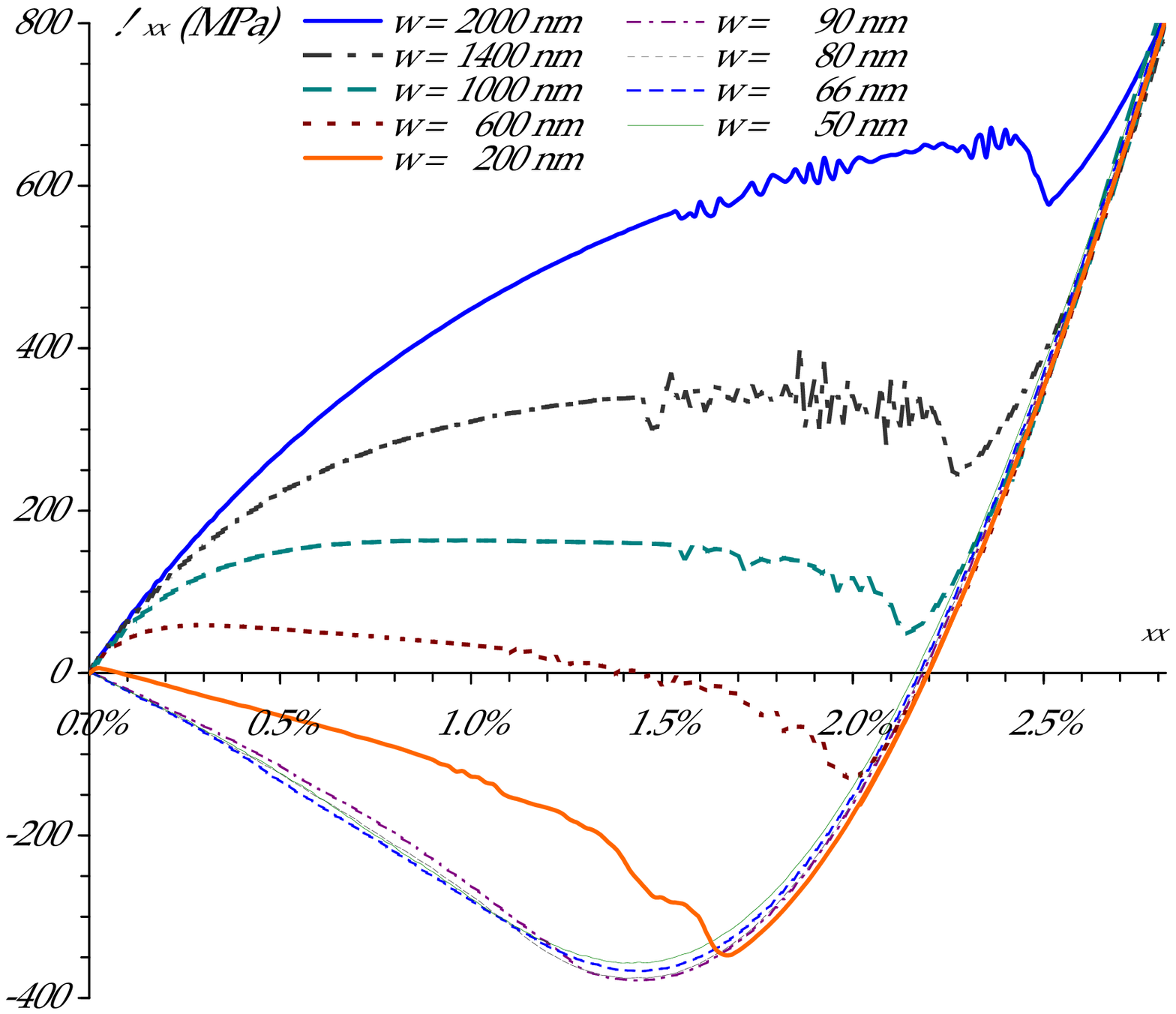}
	\caption{\label{ss}Stress--strain curves (loading only) for several nanowire widths. Microstructures are shown in Figs.~\ref{mech-1000_0} and~\ref{mech-50_0}--\ref{mech-200_0}.}
\end{figure}

\begin{figure}
	\centering
    \includegraphics[width=8.6cm]{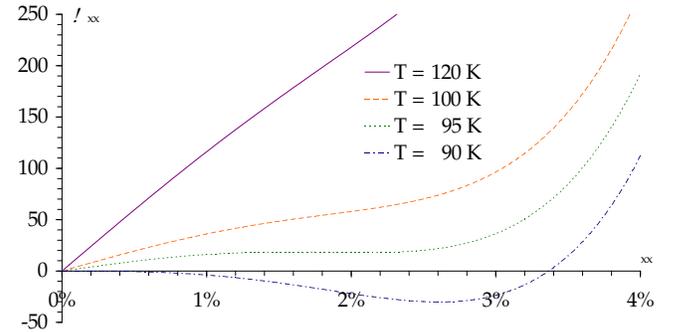}
	\caption{\label{theoretical-temp}Analytical stress--strain curves for different temperatures.}
\end{figure}

\subsection{Effect of nanowire width}
Figure~\ref{theoretical} shows that the behavior of a nanowire (even a rather wide one) is different from the bulk. One can expect that narrower nanowires will be even more different.
Figure~\ref{ss} shows stress--strain curves for several nanowire widths.\footnote{Nanowires of width 1400~nm are 1000~nm long instead of 2~$\mu$m and wires of width 2000~nm are 500~nm long. Given that these systems have very neatly periodic twins, such smaller systems are not expected to behave noticeably differently (there are periodic boundaries along the nanowire axis in either case). Simulations for $w=1000$~nm show the difference between lengths of 1000 and 2000~nm to be about 10~MPa.} Notice that as the width is decreased, the martensite yield stress decreases, i.e.\ it becomes easier to move the twin boundaries. It is instructive to compare this behavior to the temperature dependence of the stress--strain curves shown in Fig.~\ref{theoretical-temp}: the effect of decreasing the width shows similarities with a decrease of temperature. On the other hand, Fig.~\ref{ss} shows that the residual strain is very weakly dependent on the width. We should also point out that the stress--strain curves with negative stresses we observe are a consequence of the strain loading. In strain loading, we can place the system in mechanically unstable regions of the stress--strain curves.

\begin{figure}
\centering
\setlength{\unitlength}{1cm}
\begin{picture}(8.5,1.5)(.1,0)
\shortstack[c]{
\subfigure{
    \label{mech-50_0-t0020}
    \includegraphics[width=8.5cm]{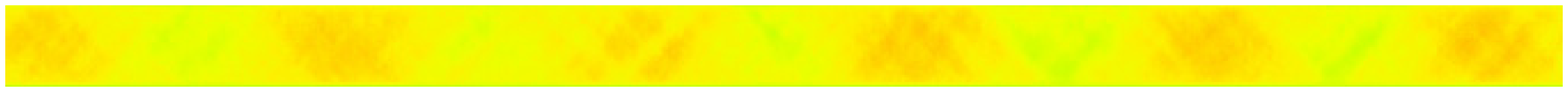}
	\put(-0.6, .15){\bf(a)}
}\vspace{-3.5mm}\\
\subfigure{
    \label{mech-50_0-t0035}
    \includegraphics[width=8.5cm]{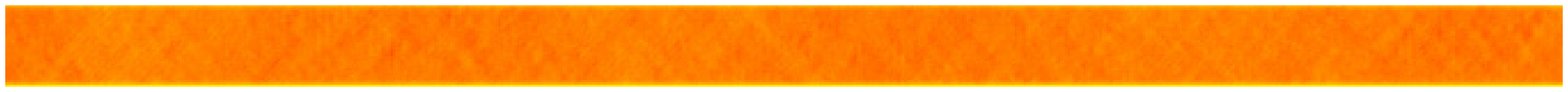}
	\put(-0.6, .15){\bf(b)}
}\vspace{-3.5mm}\\
\subfigure{
    \label{mech-50_0-t0140}
    \includegraphics[width=8.5cm]{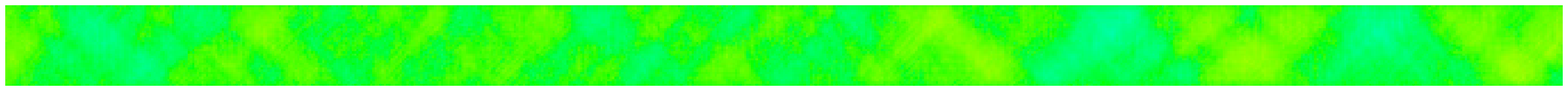}
	\put(-0.6, .15){\bf(c)}
}}
\end{picture}
	\caption{\label{mech-50_0}(Color online) Microstructure of a nanowire of width $w=50$~nm. 
(a)~$t=20$, $\varepsilon = 1\%$; 
(b)~$t=35$, $\varepsilon = 1.75\%$; and 
(c)~$t=140$, $\varepsilon = 0\%$.}
\end{figure}

One can suspect that these differences in the mechanical behavior spring from microstructural differences (we did observe in the previous section that the microstructure was strongly affected by the width of the nanowire).
Figure~\ref{mech-50_0} shows the microstructure evolution for a narrow wire ($w=50$~nm). In Fig.~\ref{mech-50_0} (and in other figures, unless otherwise specified) only half of the length of the nanowire is shown; austenite is in green and the two variants of martensite are in red and blue. Since such a wire does not form martensite in the absence of strain [see Fig.~\ref{diag-wire}], at the beginning of the tension test the wire is completely austenitic. The system is more or less uniform throughout the transformation~--- no twins or dots form. The contrast observed in Fig.~\ref{mech-50_0-t0020} is rather weak (compare to Fig.~\ref{mech-1000_0}) and is closer to the growth of an instability than to the nucleation of martensite.

\begin{figure}
\centering
\setlength{\unitlength}{1cm}
\begin{picture}(8.5,2.6)(.1,0)
\shortstack[c]{
\subfigure{
    \label{mech-66_0-t0010}
    \includegraphics[width=8.5cm]{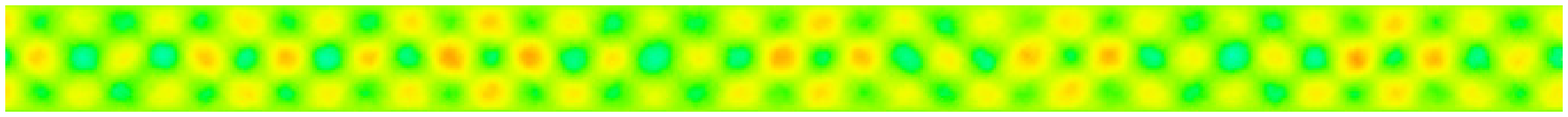}
	\put(-0.6, .25){\bf(a)}
}\vspace{-3.5mm}\\
\subfigure{
    \label{mech-66_0-t0015}
    \includegraphics[width=8.5cm]{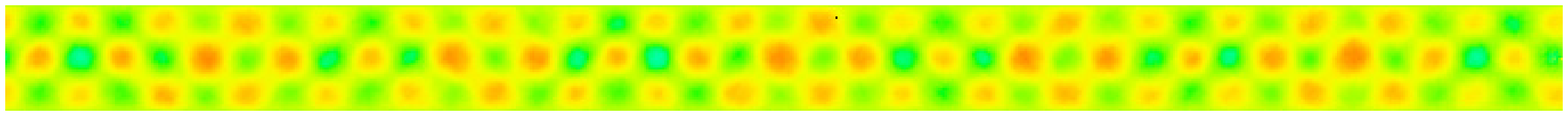}
	\put(-0.6, .25){\bf(b)}
}\vspace{-3.5mm}\\
\subfigure{
    \label{mech-66_0-t0035}
    \includegraphics[width=8.5cm]{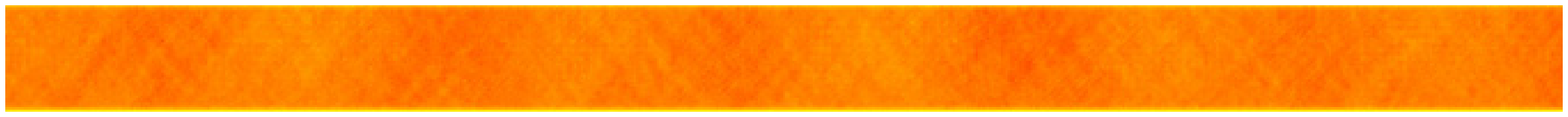}
	\put(-0.6, .25){\bf(c)}
}\vspace{-3.5mm}\\
\subfigure{
    \label{mech-66_0-t0140}
    \includegraphics[width=8.5cm]{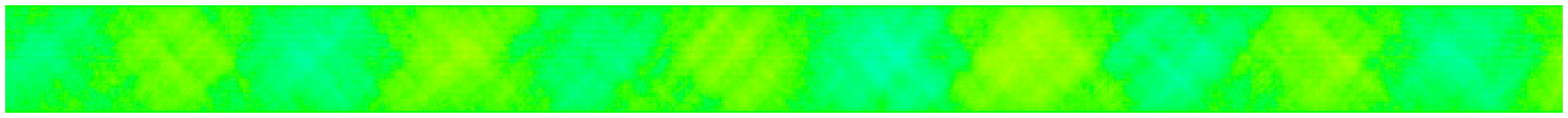}
	\put(-0.6, .25){\bf(d)}
}}
\end{picture}
	\caption{\label{mech-66_0}(Color online) Microstructure of a nanowire of width $w=66$~nm. 
(a)~$t=10$,		$\varepsilon = 0.5\%$; 
(b)~$t=15$,		$\varepsilon = 0.75\%$; 
(c)~$t=35$,		$\varepsilon = 1.75\%$; 
(d)~$t=140$,	$\varepsilon = 0\%$.} 
\end{figure}

Figure~\ref{mech-66_0} shows the time evolution of the microstructure of a wire of width $w=66$~nm, i.e.\ one of the widest wires that is austenitic at equilibrium. Figures~\ref{mech-66_0-t0010} and~\ref{mech-66_0-t0015} show that martensite appears in the form of dots of the favored variant (shown in red in Fig.~\ref{mech-66_0}). Note that this is different from the microstructure of Fig.~\ref{wire-72_0}, as it is made of one martensite variant and austenite, rather than of the two martensite variants. The austenite then disappears completely and the system in completely martensitic, Fig.~\ref{mech-66_0-t0035}. 

\begin{figure}
\centering
\setlength{\unitlength}{1cm}
\begin{picture}(8.5,2.3)(.1,0)
\shortstack[c]{
\subfigure{
    \label{mech-80_0-t0000}
    \includegraphics[width=8.5cm]{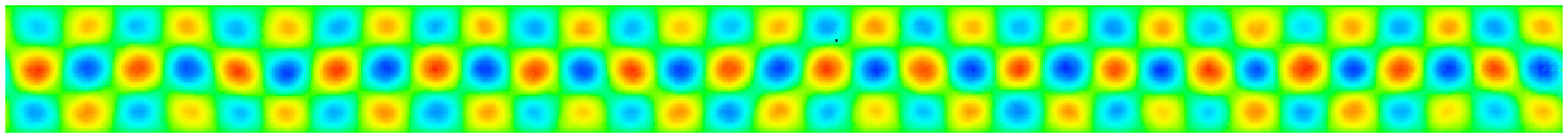}
	\put(-0.6, .35){\bf(a)}
}\vspace{-3.5mm}\\
\subfigure{
    \label{mech-80_0-t0010}
    \includegraphics[width=8.5cm]{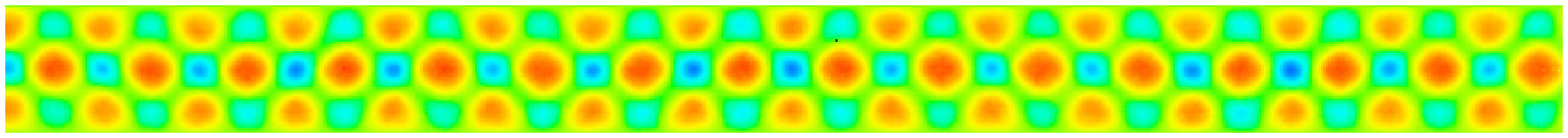}
	\put(-0.6, .35){\bf(b)}
}\vspace{-3.5mm}\\
\subfigure{
    \label{mech-80_0-t0140}
    \includegraphics[width=8.5cm]{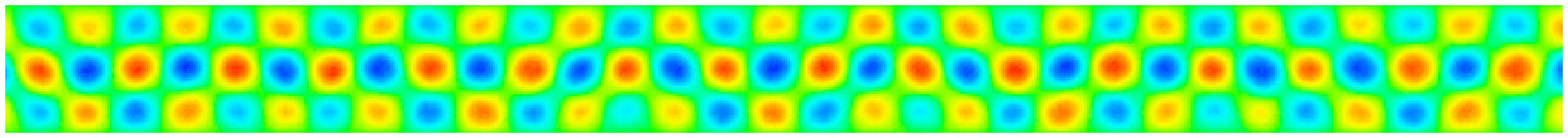}
	\put(-0.6, .35){\bf(c)}
}}
\end{picture}
	\caption{\label{mech-80_0}(Color online) Microstructure of a nanowire of width $w=80$~nm. 
(a)~$t=0$,		$\varepsilon = 0\%$; 
(b)~$t=10$,		$\varepsilon = 0.5\%$; and
(c)~$t=140$,	$\varepsilon = 0\%$.}
\end{figure}

Figure~\ref{mech-80_0} shows the microstructure evolution of a wire (of width $w=80$~nm) that is made of martensite dots at equilibrium, Fig.~\ref{mech-80_0-t0000}. Figure~\ref{mech-80_0-t0010} shows that upon loading the dots of the favorable variant (in red) expand while the dots of the other variant (blue) shrink or transform to austenite. Finally a single variant remains, similar to Fig.~\ref{mech-66_0-t0035}. (Wires of all widths up to about 250~nm have the same microstructure at $t=35$, i.e.\ $\varepsilon = 1.75\%$, as that shown in Fig.~\ref{mech-66_0-t0035}).

\begin{figure}
\centering
\setlength{\unitlength}{1cm}
\begin{picture}(8.5,3.4)(.1,0)
\shortstack[c]{
\subfigure{
    \label{mech-90_0-t0000}
    \includegraphics[width=8.5cm]{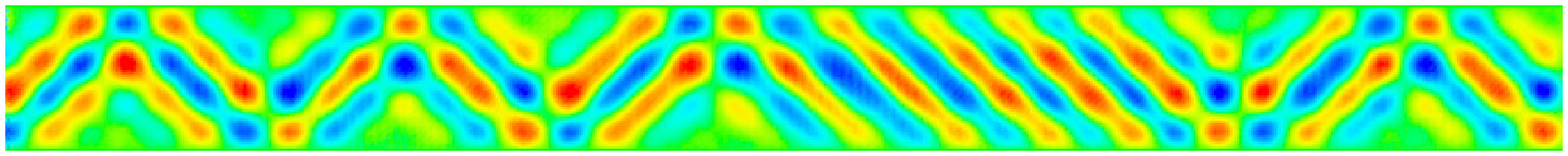}
	\put(-0.6, .4){\bf(a)}
}\vspace{-3.5mm}\\
\subfigure{
    \label{mech-90_0-t0010}
    \includegraphics[width=8.5cm]{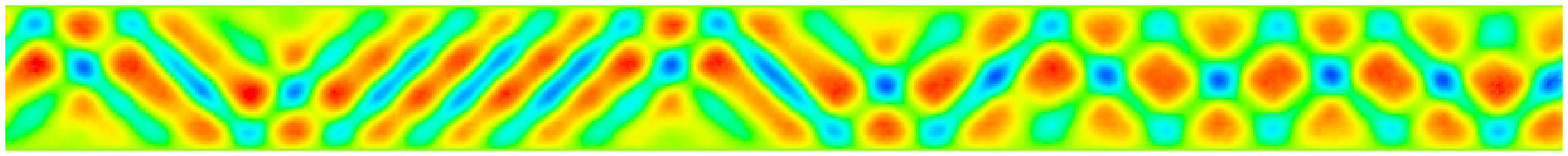}
	\put(-0.6, .4){\bf(b)}
}\vspace{-3.5mm}\\
\subfigure{
    \label{mech-90_0-t0025}
    \includegraphics[width=8.5cm]{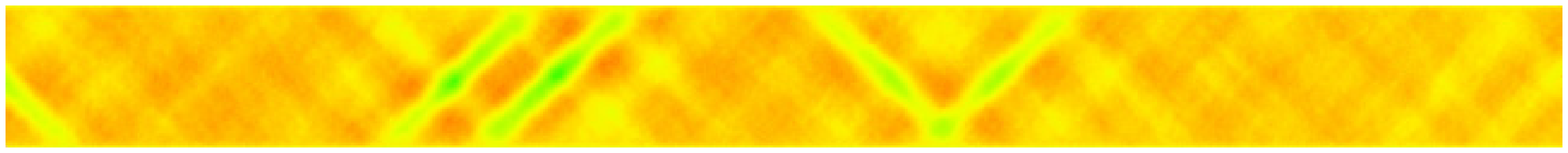}
	\put(-0.6, .4){\bf(c)}
}\vspace{-3.5mm}\\
\subfigure{
    \label{mech-90_0-t0140}
    \includegraphics[width=8.5cm]{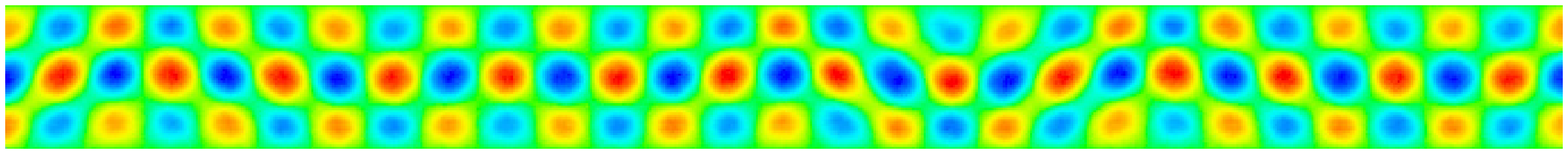}
	\put(-0.6, .4){\bf(d)}
}}
\end{picture}
	\caption{\label{mech-90_0}(Color online) Microstructure of a nanowire of width $w=90$~nm. 
(a)~$t=0$,		$\varepsilon = 0\%$;
(b)~$t=10$,		$\varepsilon = 0.5\%$;
(c)~$t=25$,		$\varepsilon = 1.25\%$; and
(d)~$t=140$,	$\varepsilon = 0\%$.} 
\end{figure}

Figure~\ref{mech-90_0} shows the microstructure evolution of a wire of width $w=90$~nm; this is the narrowest wire  exhibiting twins at equilibrium. As expected the twins corresponding to the favored variant grow and the unfavorable variant shrinks. Figure~\ref{mech-90_0-t0010} also shows a change in microstructure in some parts of the wire: some of the twins turn into square dots with faces along $\langle 11 \rangle$. 

Even though the microstructure evolution is quite different between $w=50$~nm (initially austenitic, no pattern formation --- Fig.~\ref{mech-50_0}), $w=66$~nm (initially austenitic, forms dots upon loading --- Fig.~\ref{mech-66_0}), $w=80$~nm (initially made of martensite dots, Fig.~\ref{mech-80_0}), and $w=90$~nm (initially twinned, Fig.~\ref{mech-90_0}), Fig.~\ref{ss} shows that there is barely any difference in the mechanical response. The mechanical behavior of these narrow wires is not directly influenced by whether there is martensite initially or the way martensite forms, i.e.\ the mechanical properties of these narrow wires are independent of the microstructure. (Note that the stress always increases initially, before decreasing; even though this linear increase may be small it always exists.)

\subsection{Unloading}
So far we focused on the loading of the nanowires. There is a microstructure difference upon unloading too: when the strain is back to zero (at $t=140$), the wires of widths $w=50$~nm and 66~nm are austenitic, Figs.~\ref{mech-50_0-t0140} and~\ref{mech-66_0-t0140}, while wider wires exhibits martensite dots, Fig.~\ref{mech-80_0-t0140}, or twins, Fig.~\ref{mech-1000_0-t0140}. 
Wires of width $w=90$~nm exhibit a microstructure after unloading that is different from the initial configuration, compare Fig.~\ref{mech-90_0-t0140} to Fig.~\ref{mech-90_0-t0000}. 

\begin{figure}
\centering
    \includegraphics[width=8.5cm]{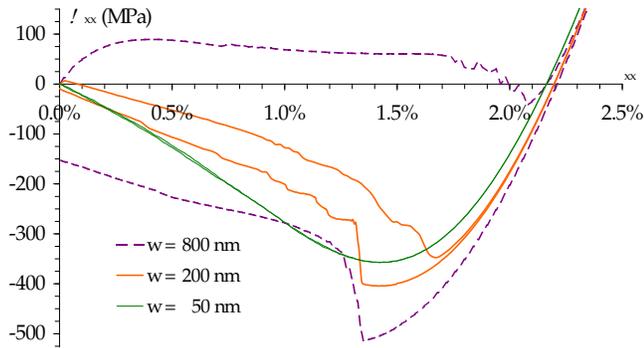}
	\caption{\label{ss-unloading}Stress--strain curves for several nanowire widths. For each width, the top curve corresponds to loading and the bottom to unloading.}
\end{figure}

The stress--strain curves for loading and unloading are almost identical for narrow wires; the difference between them increases as the nanowire width increases, as seen in Fig.~\ref{ss-unloading}. An important consequence of this hysteresis is that even when the stress never becomes negative on loading there can be a residual strain when one unloads. Yet, one expects that wide enough wires should behave like the bulk, i.e.\ pseudoelastically. There must therefore be a transition from shape-memory to pseudoelastic at some sufficient width. (Since we did not observe it, we can assume that the necessary width is too great to be computationally tractable.) 

\begin{figure}
\centering
\setlength{\unitlength}{1cm}
\begin{picture}(8.5,2.8)(.1,0)
\shortstack[c]{
\subfigure{
    \label{mech-200_0-t0113}
    \includegraphics[width=8.5cm]{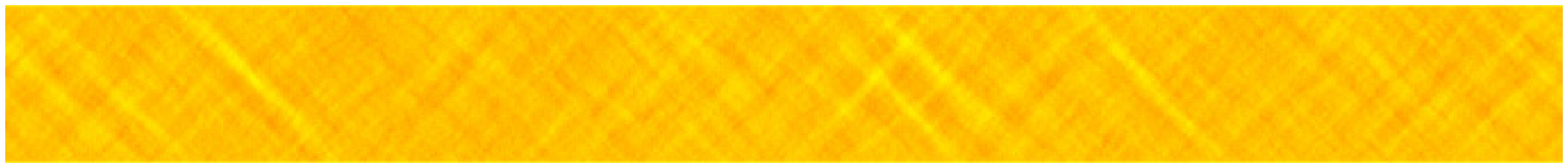}
	\put(-0.6, 0.5){\bf(a)}
}\vspace{-3.5mm}\\
\subfigure{
    \label{mech-200_0-t01132}
    \includegraphics[width=8.5cm]{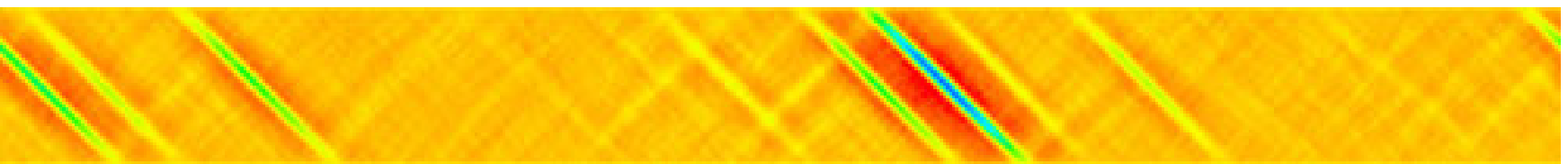}
	\put(-0.6, 0.5){\bf(b)}
}\vspace{-3.5mm}\\
\subfigure{
    \label{mech-200_0-t01136}
    \includegraphics[width=8.5cm]{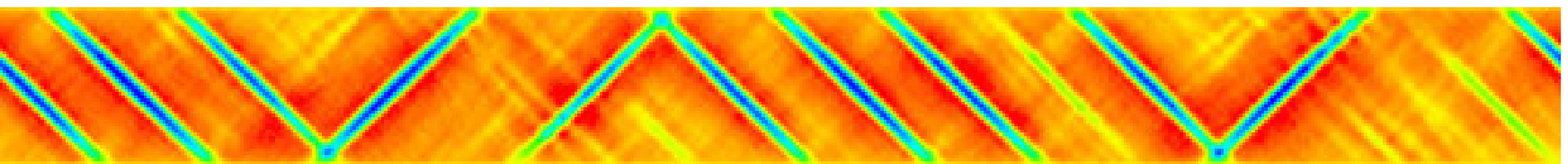}
	\put(-0.6, 0.5){\bf(c)}
}}
\end{picture}
	\caption{\label{mech-200_0}(Color online) Microstructure of a nanowire of width $w=200$~nm. 
(a)~$t=113$,	$\varepsilon = 1.35\%$; 
(b)~$t=113.2$,	$\varepsilon = 1.34\%$; and
(c)~$t=113.6$,	$\varepsilon = 1.32\%$.
(Unlike in other figures the whole system, rather than only one half, is shown.)}
\end{figure}

The sharp increase of stress visible in Fig.~\ref{ss-unloading} during the unloading of the wire of width $w=200$~nm occurs between $t \approx 113.2$ [$\varepsilon = 1.34\%$, Fig.~\ref{mech-200_0-t01132}] and $t \approx 113.6$ [$\varepsilon = 1.32\%$, Fig.~\ref{mech-200_0-t01136}]. It coincides with the nucleation and growth of the other variant (in blue): this variant is visible in Figs.~\ref{mech-200_0-t01132} and~\ref{mech-200_0-t01136} but absent from Fig.~\ref{mech-200_0-t0113}.

\subsection{\label{sec-stress_max}Martensite yield stress}
\begin{figure}
\centering
    \includegraphics[width=8.6cm]{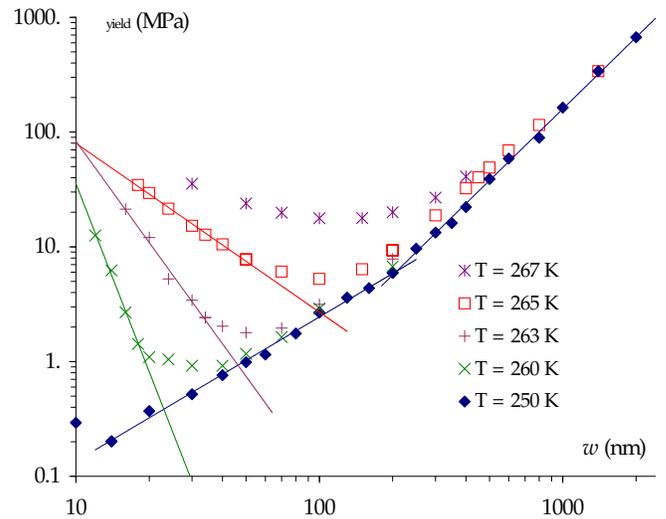}
	\caption{\label{stress_max}The martensite yield stress as a function of the width $w$ of the nanowire, for different temperatures.}
\end{figure}

Figure~\ref{stress_max} shows, as a function of the width $w$, the value of the yield stress (i.e.\ the maximum value of the stress after the initial linear part in Fig.~\ref{ss}). One can see in Fig.~\ref{stress_max} that the maximum value reached by the stress upon loading increases with the wire width (closed symbols). There are several regimes: for widths lower than about 200~nm, $\sigma_\mathrm{yield}$ has a power law dependence on $w$ with a power of 1.25, whereas the power is 2 for wider wires. For very narrow wires, however, the yield stress gets larger. 

\begin{figure}
	\centering
    \includegraphics[width=8.5cm]{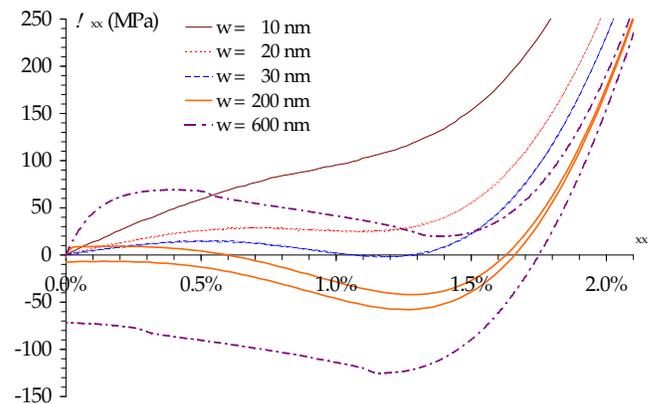}
	\caption{\label{ss-265K}Stress--strain curves for nanowires of various widths at $T=265$~K and $x_{12}=0$.}
\end{figure}

Figure~\ref{ss-265K} shows stress--strain curves for $T=265$~K (i.e.\ the transition temperature). At this temperature, curves are shifted toward lower stresses for increasing thickness until about 100~nm. For wider wires the stress increases with thickness. Very narrow wires ($w<30$~nm) behave pseudoelastically, whereas wider nanowires are in the shape-memory regime. (At $T=250$~K, all nanowires are in the shape-memory regime.) 
Figure~\ref{stress_max} shows the 
yield stress as a function of the nanowire width. For $T=265$~K (open symbols), the yield stress decreases then increases with nanowire width, with a minimum around $w=100$~nm. Wires narrower than about 18~nm do not exhibit a yield stress: the stress increases monotonically with the strain, see for instance $w=10$~nm in Fig.~\ref{ss-265K}.
The yield stress obeys a power law of exponent $-1.5$ for narrow wires ($w \leq 50$~nm) 
and another power law for wide wires (if $w \geq 200$~nm, the exponent is 1.8). 
For higher temperatures, such as $T=260$~K and $T=263$~K, there are also two regimes, but the minimum is reached at lower and lower width as the temperature decreases. On the other hand, for the wider wires the yield stress is essentially the same at all five temperatures.

\subsection{Going to a single variant}
\begin{figure}
	\centering
    \includegraphics[width=8.6cm]{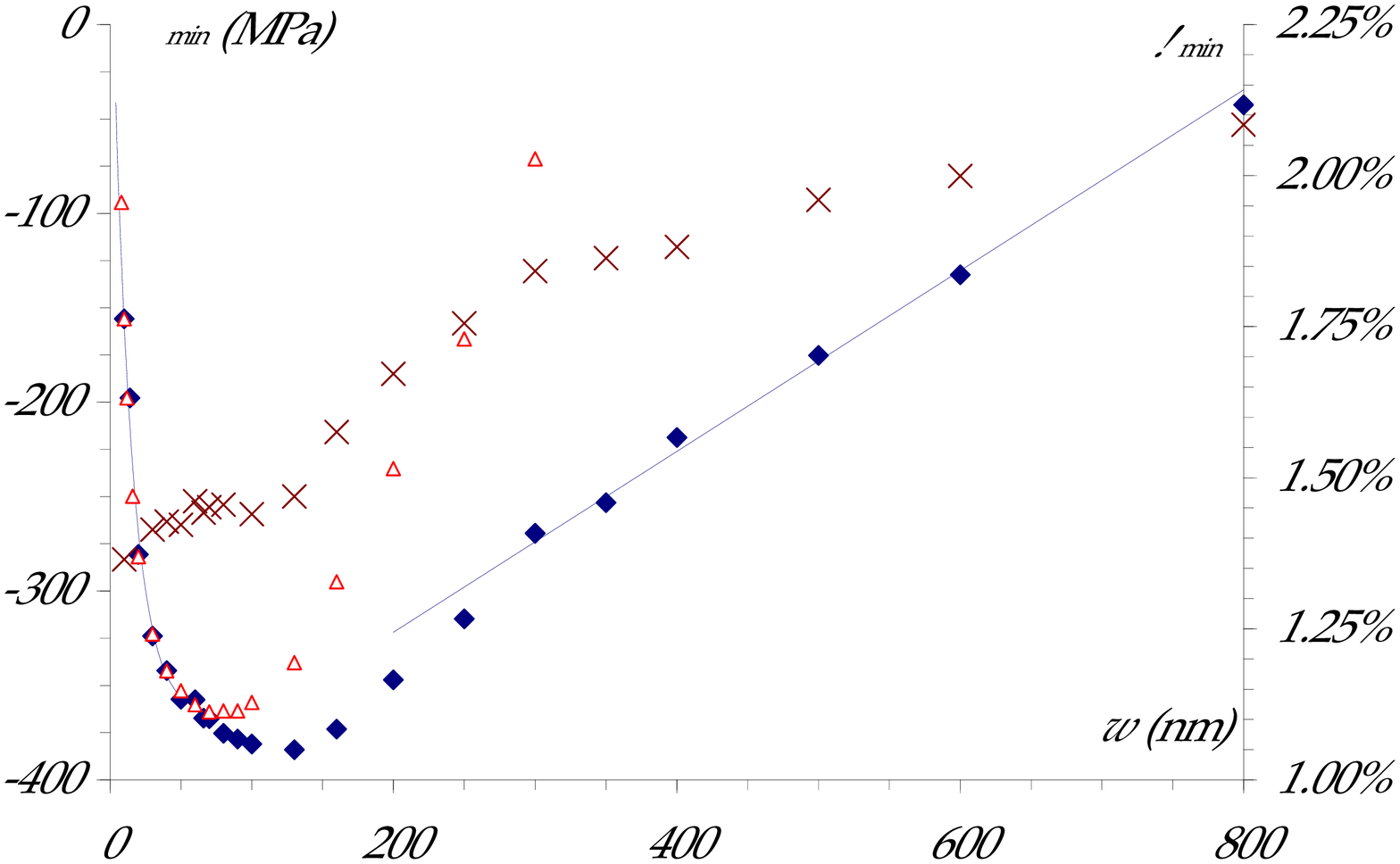}
	\caption{\label{stress_min}The minimum value of the stress $\sigma_\mathrm{xx}$ in the stress--strain curve of Fig.~\ref{ss} as a function of the width $w$ of the nanowire. 
Crosses are the strains at which the stress is minimum and empty triangles correspond to $\sigma_\mathrm{min}$ for a faster loading of $\dot{\varepsilon} = 0.5$\% per unit time. 
}
\end{figure}

Figure~\ref{ss} shows that the minimum value reached by the stress $\sigma_\mathrm{xx}$ upon loading and the strain at which this minimum is reached vary little for narrow wires but increase for wider wires.
Figure~\ref{stress_min} shows these as a function of the nanowire width $w$. There are three regimes for the stress: for narrow wires ($w \leq 50$~nm) $\sigma_\mathrm{min} \approx 439.2 \exp(-w/13.4) -367.5$; for wires wider than 300~nm, $\sigma_\mathrm{min}$ increases with the width linearly ($\sigma_\mathrm{min} \approx 0.45\,w - 405$); the smallest value of $\sigma_\mathrm{min}$ is obtained in the transition regime in-between, for nanowires 130~nm wide. (The linear fit of Fig.~\ref{stress_min} is based on data for nanowire widths up to $w=2000$~nm.)

\begin{figure}
	\centering
    \includegraphics[width=8.5cm]{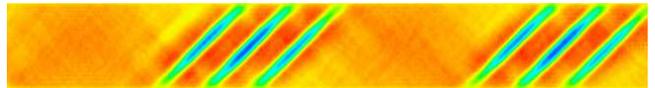}
	\caption{\label{mech-130_0-t0025}(Color online) Microstructure of a nanowire of width $w=130$~nm at $t=25$ ($\varepsilon = 1.25\%$).}
\end{figure}

The strain at which the stress is minimum (shown as crosses in Fig.~\ref{stress_min}) also exhibits several regimes: for nanowires narrower than about 130~nm, this strain does not depend on width and it increases with $w$ for wider wires (above 300~nm, the dependence on width is weaker than between 130~nm and 300~nm).
Figure~\ref{mech-130_0-t0025} shows that a nanowire of width $w=130$~nm exhibits clear twins of the unfavored variant (blue), whereas a nanowire of width $w=90$~nm does not, Fig.~\ref{mech-90_0-t0025}. The change of regime around $w=130$~nm seems to come from these unfavorable twins, which require higher and higher strains to disappear when the thickness of the wire increases.

\section{Effects of volume change and strain rate}
All tension tests presented in the previous section were carried for a system that does not have a volume change, i.e.\ $x_{12}=0$ in Eq.~(\ref{eq-G}), and for a strain rate of 0.05\% per time unit. In the present section, we will study the consequences for microstructures and mechanical properties of nanowires of varying these parameters. All simulations are at $T=250$~K.

\begin{figure}
	\centering
    \includegraphics[width=8.5cm]{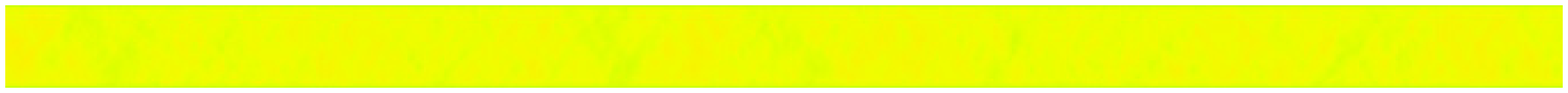}
	\caption{\label{mech-50_10-t0020}(Color online) Microstructure of a nanowire of width $w=50$~nm for a volume change $x_{12}=10$ at $t=20$ ($\varepsilon = 1\%$).}
\end{figure}

\subsection{Volume change}

\begin{figure}
	\centering
	\includegraphics[width=8.5cm]{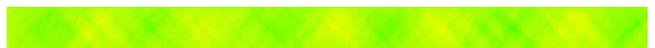}
	\caption{\label{mech-66_10-t0010}(Color online) Microstructure of a nanowire of width $w=66$~nm for $x_{12}=10$ at $t=10$ ($\varepsilon = 0.5\%$).}
\end{figure}

\begin{figure}
\centering
\setlength{\unitlength}{1cm}
\begin{picture}(8.5,3.1)(.1,0)
\shortstack[c]{
\subfigure{
    \label{mech-80_10-t0000}
    \includegraphics[width=8.5cm]{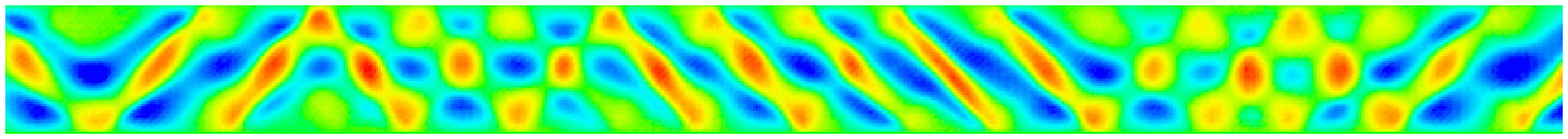}
	\put(-0.6, .4){\bf(a)}
}\vspace{-3.5mm}\\
\subfigure{
    \label{mech-80_10-t0010}
    \includegraphics[width=8.5cm]{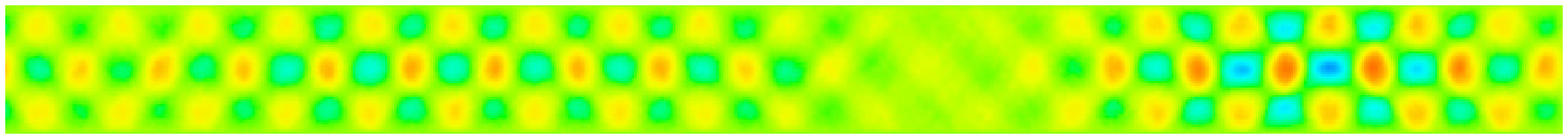}
	\put(-0.6, .4){\bf(b)}
}\vspace{-3.5mm}\\
\subfigure{
    \label{mech-80_10-t0020}
    \includegraphics[width=8.5cm]{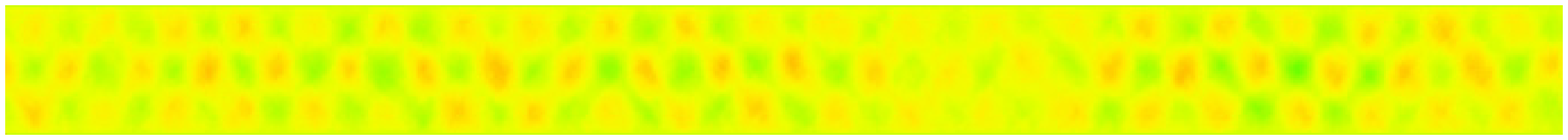}
	\put(-0.6, .4){\bf(c)}
}\vspace{-3.5mm}\\
\subfigure{
    \label{mech-80_10-t0035}
    \includegraphics[width=8.5cm]{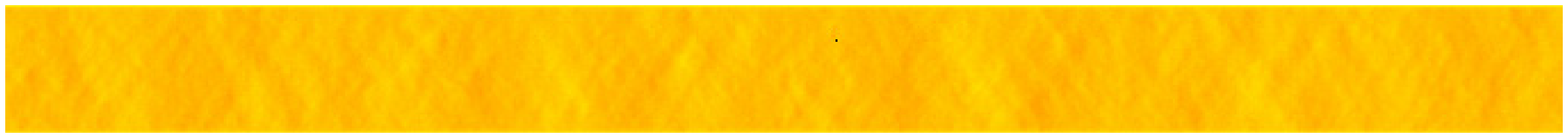}
	\put(-0.6, .4){\bf(d)}
}}
\end{picture}
	\caption{\label{mech-80_10}(Color online) Microstructure of a nanowire of width $w=80$~nm ($x_{12}=10$). 
(a)~$t=0$,		$\varepsilon = 0\%$; 
(b)~$t=10$,		$\varepsilon = 0.5\%$; 
(c)~$t=20$,		$\varepsilon = 1\%$; and
(d)~$t=35$,		$\varepsilon = 1.75\%$.}
\end{figure}

\begin{figure}
\centering
\setlength{\unitlength}{1cm}
\begin{picture}(8.5,1.7)(.1,0)
\shortstack[c]{
\subfigure{
    \label{mech-90_10-t0000}
    \includegraphics[width=8.5cm]{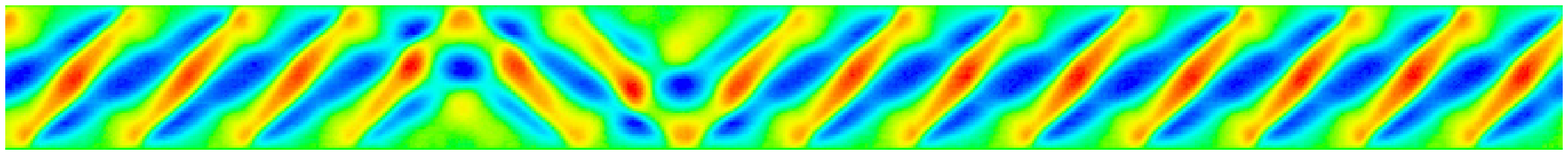}
	\put(-0.6, .45){\white(a)}
}\vspace{-3.5mm}\\
\subfigure{
    \label{mech-90_10-t0010}
    \includegraphics[width=8.5cm]{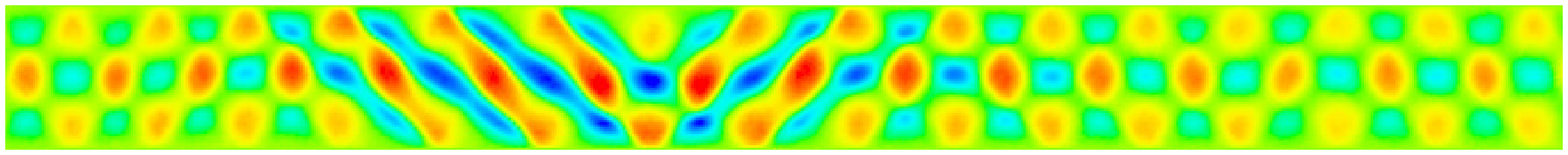}
	\put(-0.6, .45){\bf(b)}
}}
\end{picture}
	\caption{\label{mech-90_10}(Color online) Microstructure of a nanowire of width $w=90$~nm ($x_{12}=10$). 
(a)~$t=0$,		$\varepsilon = 0\%$ and 
(b)~$t=10$,		$\varepsilon = 0.5\%$.}
\end{figure}

\begin{figure}
\centering
\setlength{\unitlength}{1cm}
\begin{picture}(8.5, 1.8)(.1,0)
\shortstack[c]{
\subfigure{
    \label{mech-200_10-t0000}
    \includegraphics[width=8.5cm]{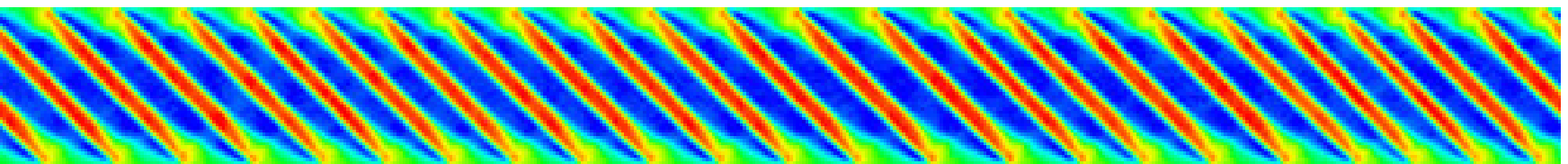}
	\put(-0.6, .5){\white(a)}
}\vspace{-3.5mm}\\
\subfigure{
    \label{mech-200_10-t0020}
    \includegraphics[width=8.5cm]{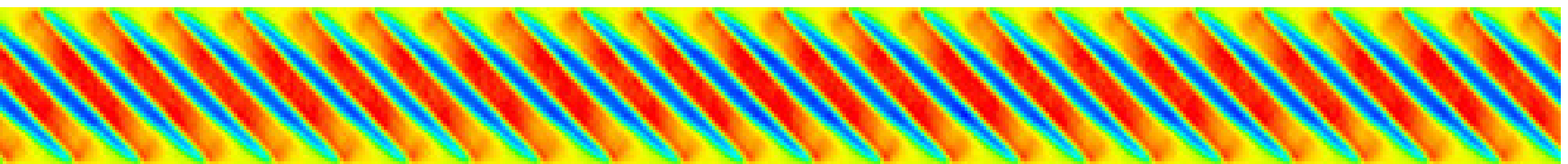}
	\put(-0.6, .5){\bf(b)}
}}
\end{picture}
	\caption{\label{mech-200_10}(Color online) Microstructure of a nanowire of width $w=200$~nm ($x_{12}=10$). 
(a)~$t=0$,		$\varepsilon = 0\%$ and 
(b)~$t=20$,		$\varepsilon = 1\%$ 
(Unlike in other figures the whole system, rather than only one half, is shown.)}
\end{figure}

The presence of a volume change (we will focus on the case of $x_{12}=10$) has little effect on the microstructure of narrow nanowires ($w=50$~nm), compare Figs.~\ref{mech-50_10-t0020} and~\ref{mech-50_0-t0020}. For nanowires of width $w=66$~nm, the volume change noticeably affects the microstructure: while the system forms martensite dots for $x_{12}=0$, Fig.~\ref{mech-66_0-t0010}, there are none if $x_{12}=10$, Fig.~\ref{mech-66_10-t0010}.
For wider wires ($w=80$ or 90~nm), the microstructure evolution in the presence of a volume change is also different from what is observed when $x_{12}=0$, see Figs.~\ref{mech-80_10} and~\ref{mech-90_10}. While in Figs.~\ref{mech-80_0} and~\ref{mech-90_0} the favored martensite variant (in red) grows at the expense of the other variant (in blue), for $x_{12}=10$ the unfavorable variant disappears almost completely but the favorable one shrinks too, this is especially noticeable in Fig.~\ref{mech-80_10-t0010}. In other words the system evolves from both martensite variants to a single variant by first turning into a mixture of austenite and (the favored variant of) martensite.
For even wider wires ($w=200$~nm), the microstructure evolution is similar to what is observed for $x_{12}=0$: compare Figs.~\ref{mech-200_10} and~\ref{mech-200_0}.

\begin{figure}
\centering
    \includegraphics[width=8.5cm]{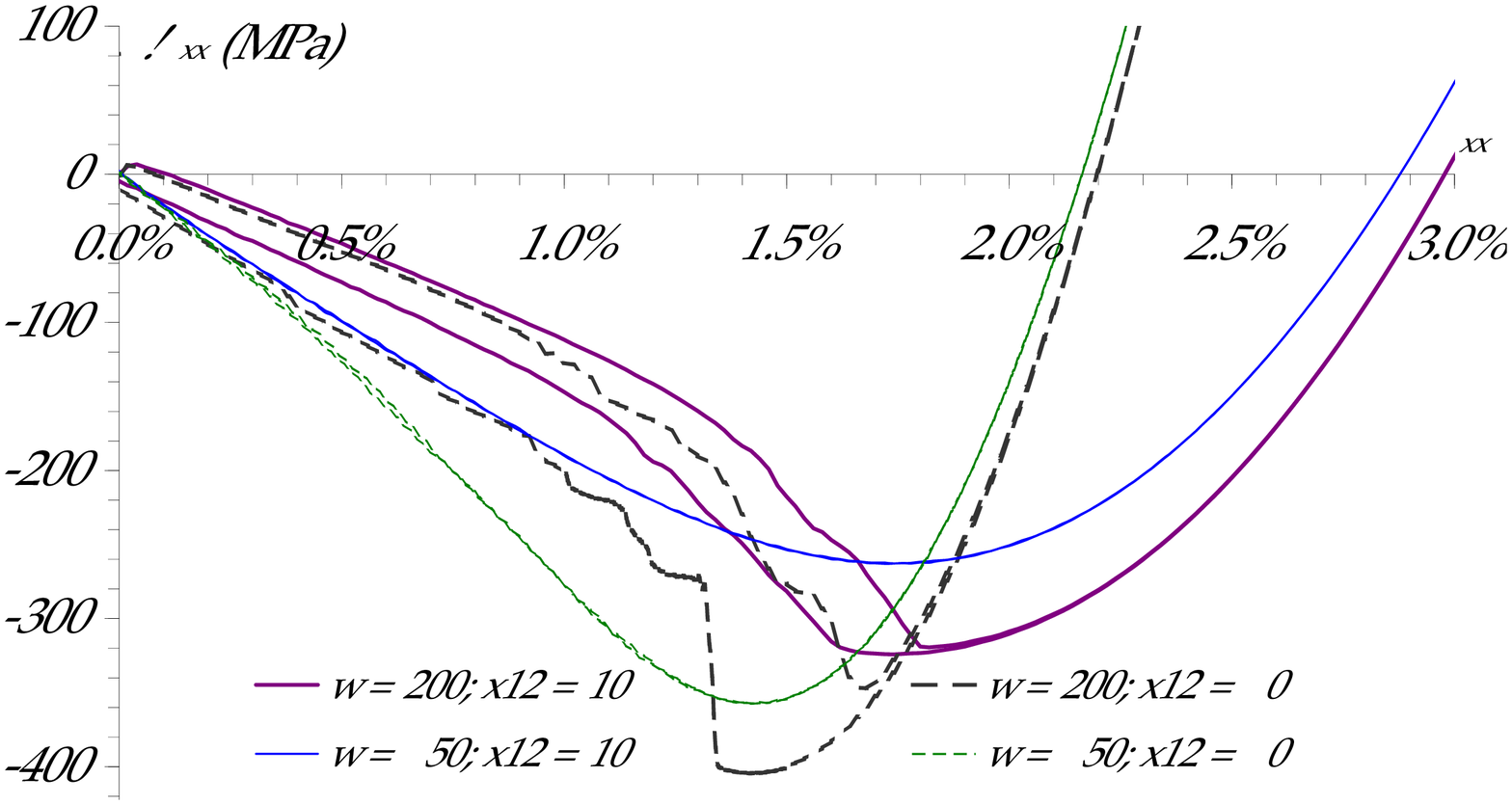}
	\caption{\label{ss-0_10}Stress--strain curves for two nanowire widths and two values of $x_{12}$.}
\end{figure}

Figure~\ref{ss-0_10} shows stress--strain curves with and without volume change. When $x_{12}=10$, the minimum stress is lower in magnitude and reached at a higher value of the strain. One can note that for $w=200$~nm there is no sudden increase of stress upon unloading.

\begin{figure}
	\centering
    \includegraphics[width=8.5cm]{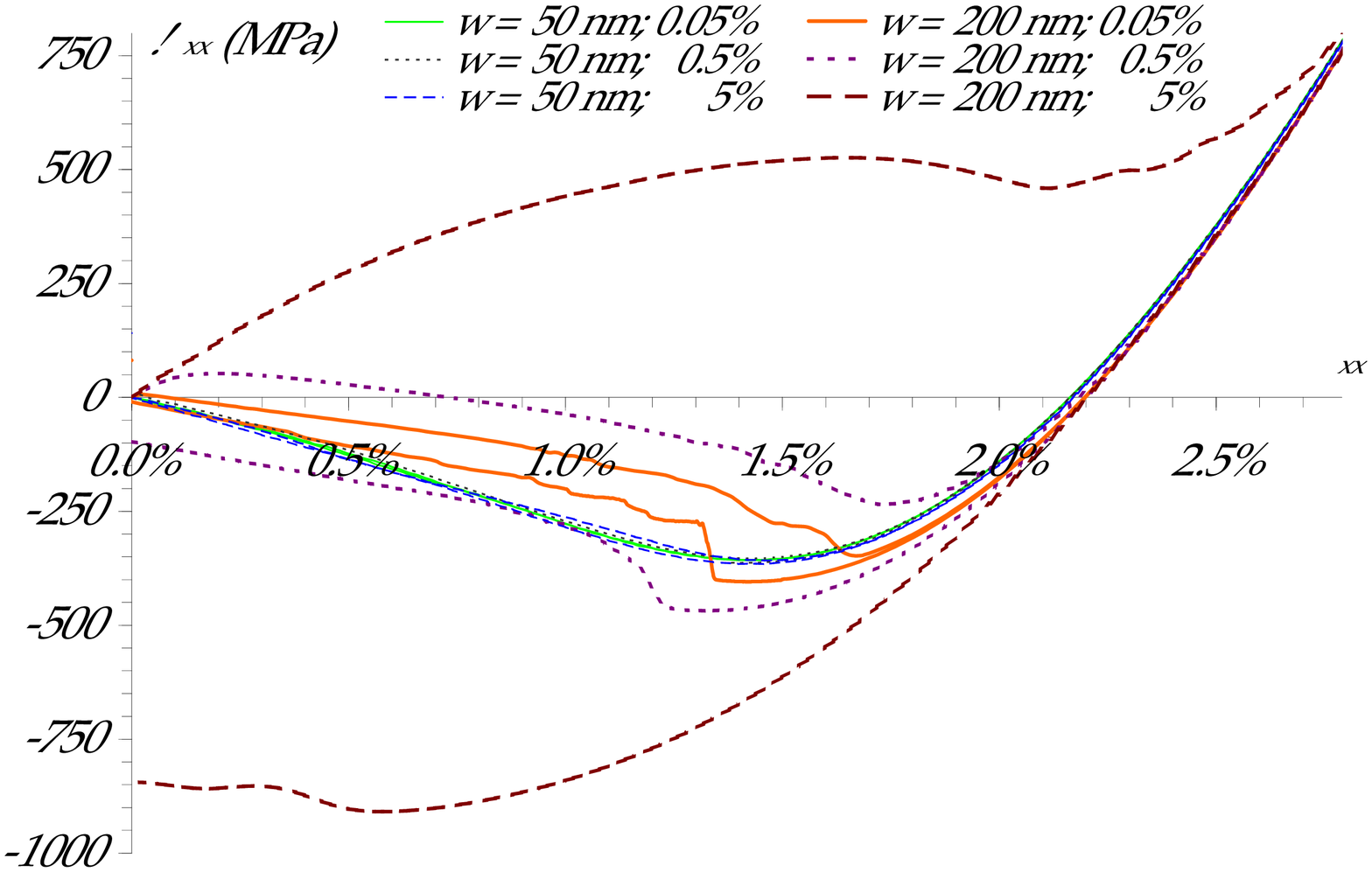}
	\caption{\label{ss-strain_rate}Stress--strain curves for nanowires of widths $w=50$~nm and $w=200$~nm at strain rates of 0.05\%, 0.5\%, and 5\% per time unit.}
\end{figure}

\subsection{Strain rate}
Figure~\ref{ss-strain_rate} shows that the stress--strain curves are affected by the strain rate. Higher strain rates shift the curve towards higher stress for loading and lower stress for unloading; consequently, the hysteresis increases with the strain rate. Since faster loading does not give twin boundaries time to move part of the strain is accommodated elastically rather than through phase transformation, which raises the stress. This effect is greater for wider wires. 

\begin{figure}
\centering
    \includegraphics[width=8.6cm]{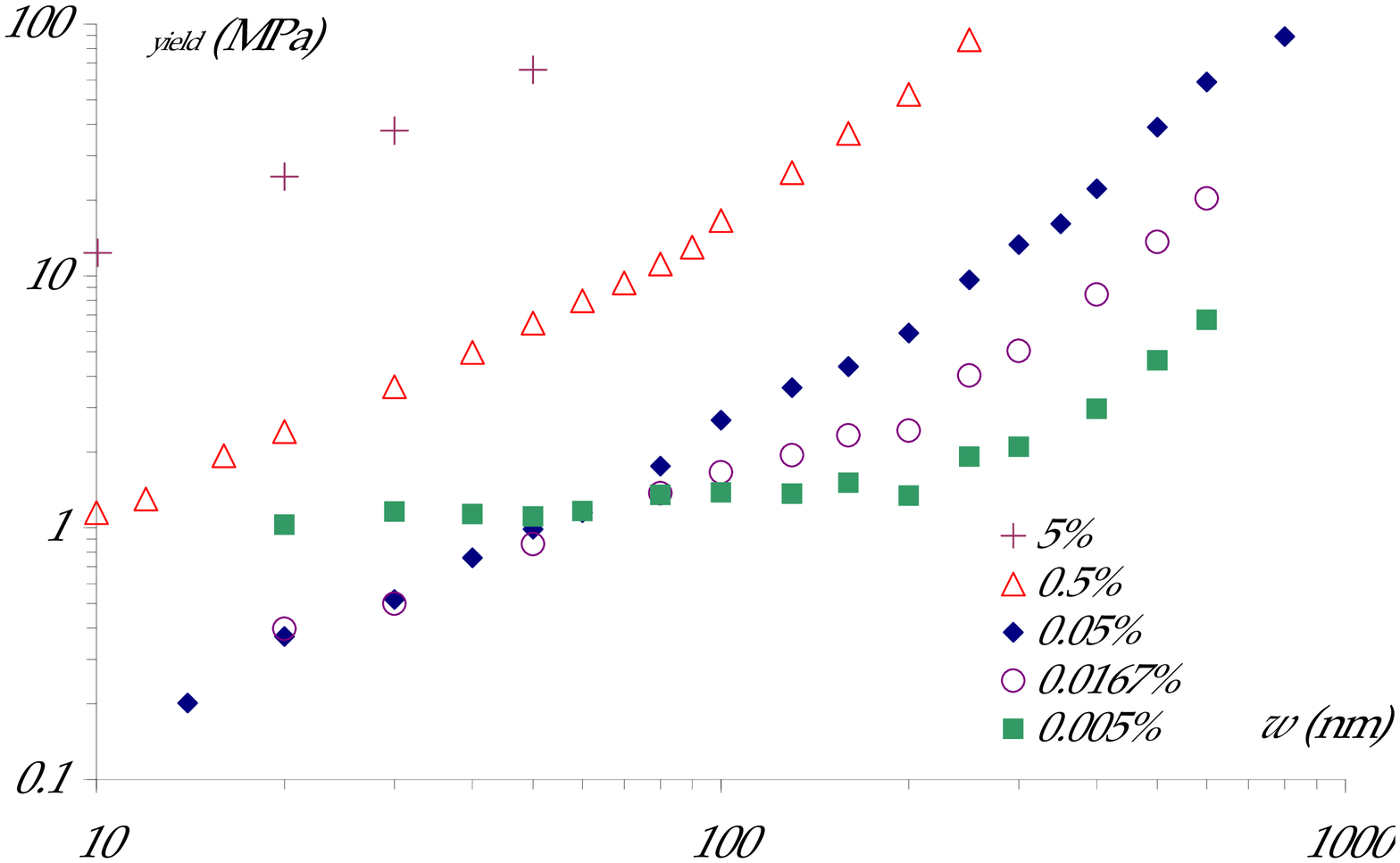}
	\caption{\label{stress_max-strain_rate}The maximum value of the stress $\sigma_\mathrm{xx}$ in the stress--strain curve of Fig.~\ref{ss} as a function of the width $w$ of the nanowire for several strain rates.}
\end{figure}

Figure~\ref{stress_max-strain_rate} shows that for faster loading the value of $\sigma_\mathrm{yield}$ increases for all widths, but this effect is more noticeable for wider wires. On the other hand, in Fig.~\ref{stress_min} $\sigma_\mathrm{min}$ does not depend on the strain rate for narrow wires (up to about $w=70$~nm).

\begin{table}
\begin{tabular}{|c|c|c|}
\hline $\sigma_\mathrm{yield}$
& slow 								& fast
\\\hline\raisebox{-1.5ex}[0pt][0pt]{narrow}	
& independent of $w$ 						& $\propto w$ \\
& $\propto \dot{\varepsilon}^{\,-1/4}$& $\propto \dot{\varepsilon}$ 
\\\hline\raisebox{-1.5ex}[0pt][0pt]{wide}
& $\propto w^n$ ($n<2$)						& $\propto w^2$ \\
& $\propto \dot{\varepsilon}$		& $\propto \dot{\varepsilon}$ 
\\\hline
\end{tabular}
\caption{Dependence of $\sigma_\mathrm{yield}$ upon nanowire width $w$ and strain rate $\dot{\varepsilon}$.}
\label{table}
\end{table}

Figure~\ref{stress_max-strain_rate} also shows that for lower strain rates (0.005\% per time unit) $\sigma_\mathrm{yield}$ does not depend on the width of the nanowire (full squares in the figure). 
Table~\ref{table} sums up the dependence of $\sigma_\mathrm{yield}$ upon the nanowire width and the strain rate.

\section{Conclusion}
We used the phase-field method to study the martensitic transformation in constrained systems at the nanoscale. For nanosystems such as nanowires and nanograins, the geometric constraints can lead to new microstructures or prevent martensite formation altogether. For all geometries, we observed a new microstructure ---made of dots aligned on a square lattice with axes along $\langle 01 \rangle$--- at intermediate size. In the case of a square grain with faces along $\langle 11 \rangle$, we also observed an asymmetric microstructure, which is not found in other geometries.

We also performed uniaxial tension tests on the nanowires. The stress--strain curves are very different from bulk results and a size dependence of mechanical properties is observed. Interestingly, the martensite yield stress shows non-monotonic behavior with respect to the nanowire width: as the width is decreased, the yield stress decreases until a critical thickness below which it starts to increase again.  At $T=265$~K, the wider wires exhibit a residual strain and are in the shape-memory regime whereas the narrower wires show pseudoelastic behavior. In contrast, at $T=250$~K a residual strain is observed for all widths. Another interesting result is that the stress--strain curves are not affected by microstructures~--- the mechanical response of systems with different microstructures may be similar, while systems with the same microstructure may have a different mechanical behavior. Finally, we also noticed that strain rate and volume change associated with the martensitic transformation have an impact of the mechanical response and on microstructures.

\bibliography{D:/work/write-ups/martensite/martensite-PRB-07/mart}

\end{document}